\documentclass{jfm}
\usepackage{graphicx}
\usepackage{epstopdf, epsfig}
\usepackage{hyperref}
\usepackage{bm}
\usepackage{color,soul}
\usepackage{mathrsfs}
\usepackage{amsmath,amssymb}

\shorttitle{Drop impact on immiscible liquid pool}

\shortauthor{D. Roy, Sophia M, and S. Basu}

\title{On the mechanics of droplet surface crater during impact on immiscible viscous liquid pool}

\author{Durbar Roy\aff{1},
Sophia M\aff{1},
\and Saptarshi Basu\aff{1}
  \corresp{\email{sbasu@iisc.ac.in}}}

\affiliation{
\aff{1}Department of Mechanical Engineering, Indian Institute of Science, Bengaluru, 560012, India
}

\begin{document}

\maketitle

\begin{abstract}
We study drop impacts on immiscible 
viscous
liquid pool and investigate the formation of droplet surface craters using experimental and theoretical analysis. 
We attribute the formation of air craters to the rapid deceleration of the droplet due to viscous drag force. The droplet response to the external impulsive decelerating force induces oscillatory modes on the surface exposed to the air forming capillary waves that superimpose to form air craters of various shapes and sizes. 
We introduce a non-dimensional parameter (${\Gamma}$), that is, the ratio of drag force to the capillary force acting on the droplet. We
show that ${\Gamma}$ is directly proportional to the capillary number. 
We show that droplets forming air craters of significant depths have ${\Gamma}>1$. Further, we demonstrate that Legendre polynomials can locally approximate the central air crater jet profile. We also decipher that the air crater response time scale ($T$) varies as the square root of impact Weber number ($T{\sim}We^{1/2}$).
Further, we generalize the local droplet response with a global response model for low-impact energies based on an eigenvalue problem.
We represent the penetrating drop as a constrained Rayleigh drop problem with a dynamic contact line. The air-water interface dynamics is analyzed using an inviscid droplet deformation model for small deformation amplitudes. The local and global droplet response theory conforms with each other and depicts that the deformation profiles could be represented as a linear superposition of eigenmodes in Legendre polynomial basis. We unearth that the droplet response in an immiscible impact system differs from the miscible impact systems due to the presence of such a dynamic contact line. 
\color{black}
\end{abstract}

\begin{keywords}
Authors should not enter keywords on the manuscript, as these must be chosen by the author during the online submission process and will then be added during the typesetting process (see http://journals.cambridge.org/data/\linebreak[3]relatedlink/jfm-\linebreak[3]keywords.pdf for the full list)
\end{keywords}

\section{Introduction}
The study of drop impact physics began with the seminal works of Worthington in the late nineteenth century \citep{worthington1877xxviii,worthington1897v} and continues till date due to applications in various industries like manufacturing, printing, food-processing, bio-medical, and pharmaceuticals \citep{bolleddula2010impact,pasandideh2002three,roy2019dynamics}. The early pioneering work of Worthington led to the discovery of various spatio-temporal physics during drop impacts on solids and liquids. Drop impacts on liquids generates aesthetically beautiful patterns like crowns, corona splash, and jets, to name a few \citep{gekle2010generation,gordillo2010generation} and are used in various artistic contexts and platforms.
The mechanism underlying the beauty is multi-scale and multi-physics in origin and has attracted many scientists and engineers to study drop impacts on liquids theoretically, numerically, and experimentally \citep{rioboo2001outcomes,sikalo2000analysis,tropea1999impact,roisman2002impact,thoraval2012karman,renardy2003pyramidal}. With the advent of ultra-high-speed imaging and laser diagnostic techniques \citep{lauterborn1984modern,adrian1991particle,thoroddsen2008high,versluis2013high} the spatio-temporal experimental resolution has increased by orders of magnitude from the early days of drop impact research \citep{worthington1877xxviii,worthington1897v}.

A droplet impacting on a liquid media can result in a wide variety of phenomena like splashing, bouncing, coalescence, and formation of jets, corollas, and crowns depending on the impact parameter space characterized by various non-dimensional groups like impact Weber number, Reynolds number, Ohnesorge number, Bond number, Froude number to name a few \citep{thoroddsen2011droplet,yarin2017collision}. The various non-dimensional numbers represent the ratios of various competing effects governing the dynamics of drop impacts. Several phenomena like the formation of flower-like pattern in the liquid film, entrapment of air, generation and propagation of ejecta sheet widen the landscape of possibilities of drop impact physics on liquids \citep{thoroddsen2002ejecta,zhang2012evolution,marcotte2019ejecta}.
During the impinging process, just prior to impact, a thin lubricating air layer gets entrapped underneath the droplet \citep{thoroddsen2003air,bartolo2006singular}.
The entrapped air between the drop and the liquid must be displaced/drained out before proper contact between the drop and the pool.
 The pressure in the lubricating air layer is inversely proportional to the air layer thickness and hence increases monotonically till the entrapped air pressure reaches the capillary pressure of the droplet 
 \citep{thoroddsen2003air,hicks2010air,hicks2011air,doi:10.1063/5.0091584}.
 The excess pressure above the capillary pressure causes a dimple to form just beneath the impacting droplet. The air layer ruptures at a point leading to the first local contact of the drop and the formation of entrapped air bubbles. The formation of air bubbles in drop impact systems can be detrimental to specific technologies such as printing, coating, and several cooling applications \citep{aziz2000impact,yarin2017collision,aksoy2020impact}. However, the entrapment of air bubbles is a boon to aquatic life forms since these entrapped air bubbles are medium of gaseous exchange between the atmosphere and the water bodies that sustain marine life forms \citep{woolf2007modelling}. For impact on similar liquids, after the initial entrapped air layer rupture phase, the droplet forms an air crater in the liquid layer/pool, collapsing to form Worthington jets. 
 The singularity dynamics and curvature collapse leading to jet erruption on a fluid surface was studied in a seminal work by Zeff et al. \citep{zeff2000singularity}.
Further, in the previous decade, 
Gekle and Gordillo studied the Worthington jet formed during the impact of a circular disc on water using detailed boundary-integral simulations and analytical modeling \citep{gekle2010generation,gordillo2010generation}. They discovered that the flow structure inside the jet could be divided into three regions; the acceleration region, ballistic region, and tip region.
Majority of the previous study on drop impact on liquids have focused on impacts on similar liquids \citep{shetabivash2014numerical,yarin2017collision,castillo2015droplet,castrejon2016droplet,hasegawa2019energy}. However, literature of drop impact on immiscible pools are relatively sparse \citep{yakhshi2010impact,dhuper2021interface,che2018impact,minami2022cavity}. Drop impact on immiscible liquid pools is very important and is found in various industrial, engineering, and natural systems. Droplet interactions in immiscible systems are inherently different, as mentioned in some of the literature available\citep{che2018impact}. In general, drop impact on liquids produces Worthington jets \citep{worthington1897v} on the liquid pool formed due to the collapse of an air crater in the liquid pool. However, we have found that the air craters formed in immiscible impact systems are significantly different from those reported in the literature for miscible liquids. Primarily, the air crater on the surface of the impacting droplet resembles central air craters found in drop impact on superhydrophobic substrates \citep{yamamoto2018initiation} at low impact Weber number. 
Locally, the 
air craters on the drop surface can be understood based on the response dynamics of the droplet subjected to external forces, as was demonstrated by Harper et al. and Simpkins et al.  \citep{harper1972breakup,simpkins1972water}.
The similarity between air craters/jets observed on droplet surface during impact on viscous immiscible liquid pools with jets/craters observed on droplet surface impacting a hydrophobic surface led us to map the immiscible liquid impact problem to the constrained Rayleigh drop eigenvalue problem \citep{strani1984free,bostwick2009capillary,bostwick2013coupled-1,bostwick2013coupled-2,bostwick2014dynamics,bostwick2015stability} as was done by Kern et al. for drop impact on solid surfaces \citep{kern2021drop}.
Recent studies have shown various topology of air crater/singular jet formation in immiscible impacts \citep{yang2020multitude}. 
However, the detailed mechanism of the jets produced on the surface of the droplets is not well understood and remains elusive.

A general three fluid immiscible system dealing with impact scenarios labelled as $1,2,3$ is described by the following physical parameters:
 the densities $({\rho}_1,{\rho}_2,{\rho}_3)$; the viscosities $({\mu}_1,{\mu}_2,{\mu}_3)$; the pair wise surface tensions $({\sigma}_{12},{\sigma}_{13},{\sigma}_{23})$; the impact velocity $V_0$; acceleration due to gravity $g$; and radius of the droplet $R_0$. Overall there are $12$ dimensional parameters to fully characterize the general dynamics of the impacting system. Using the Buckingham-Pi theorem \citep{barenblatt1996scaling} we can construct $9$ non-dimensional parameters to reduce the parameter space. The non-dimensional parameters are ${\rho}_1/{\rho}_2$, ${\rho}_2/{\rho}_3$, ${\mu}_1/{\mu}_2$, ${\mu}_2/{\mu}_3$, ${\sigma}_{12}/{\sigma}_{13}$, ${\sigma}_{13}/{\sigma}_{23}$, $We$, $Ca$, $Fr$ where $We$, $Ca$ and $Fr$ are the Weber, Capillary and Froude numbers respectively. We observe out of the $9$ non-dimensional quantities only $3$ quantities are related to dynamics and the remaining $6$ quantities specifies the fluid properties and defines the particular fluids used. For a given set of fluids, the dynamics could be characterized based on the triplets ($We$, $Ca$, $Fr$). The Weber number based on the droplet diameter is defined as $We={\rho}_wV_0^22R_0/{\sigma}_{aw}$, where ${\rho}_w$ is the density of the impacting droplet, $V_0$ is the impact velocity and ${\sigma}_{aw}$ is the air-water surface tension. The capillary number is defined as $Ca={\mu}_sV_0/{\sigma}_{aw}$, where ${\mu}_s$ is the viscosity of silicone oil pool. The Froude number is defined as $Fr=V_0/{gR_0}$. Note that the three triplets $(We,Ca,Fr)$ are monotonically increasing function of the impact velocity $V_0$. The Weber number depends on the impact velocity quadratically whereas the Capillary and Froude number varies linearly. The main experimental control parameter used in this work is the impact velocity and hence the Weber number has the largest variation in terms of numerical values. We use Weber number to characterize all the observations experimentally since an order of magnitude change is observed in We due to the variation in the range of $V_0$ . Note however for different values of Weber number, the Capillary number and the Froude number are uniquely defined and any impact event in the current experimental context of the work requires a triplet of $(We, Ca, Fr)$ to uniquely define an impact event. In general, the force involved to cause droplet retardation is viscous drag. However, the observed response of the droplet is governed by inertia and surface tension. This is due to the fact that
the impacting droplet is about $350$ times less viscous than the silicone oil pool. The experimental response time scale of the droplet is within the capillary time scale ($t_c{\sim}\mathcal{O}(9{\times}10^{-3}s)$). The viscous effects inside the droplet are important at the scale of the viscous diffusion time scale (${\sim}\mathcal{O}(4R_0^2/{\nu}_w)=5.43s$). We should realize that the droplet response occurs at a scale of milliseconds which is three orders of magnitude smaller than the viscous diffusion time scale inside the droplet. Therefore, we can use Weber number to describe the crater observations in general. Further, the high viscosity of the pool ensures that the surface of the droplet below the dynamic contact line
has negligible modal oscillations due to high dissipation. The time scale of the viscous dissipation in the liquid pool is of the order of
the air crater response time scale (${\sim}\mathcal{O}(4R_0^2/{\nu}_s){\sim}10{\times}10^{-3}s{\sim}t_c$).
In this work, we study the role of 
impact Weber number ($We={\rho}_wV_0^22R_0/{\sigma}_{aw}$) on the formation of air craters on the surface of an impacting water droplet on an immiscible 
highly viscous 
liquid pool of silicone oil with particular focus on the mechanism of air crater formation using experimental and theoretical analysis. The remaining text is organized as follows. In section 2, we provide the details of the experimental setup and data processing 
methods. 
Section 3 outlines the results and discussions with various subsections. Subsection 3.1 provides a global overview of the important results obtained. Subsection 3.2 discusses droplet penetration dynamics and the scales of the various forces acting on the droplet. Subsection 3.3 details ways of characterizing droplet deformation from the spherical shape. Subsection 3.4 provides a criterion for forming air craters of significant depth. Subsection 3.5 computes the center of mass velocity of the droplet. Section 3.6 outlines the details of droplet response/air crater and jet characteristics 
from a local analysis view point. Section 3.7 discusses the droplet response from a constrained Rayleigh drop model perspective using the Green function method to solve an eigen value problem. 
We conclude with the results 
and future scope
in conclusion (section 4). The underlying theme of the current work was to identify the mechanisms and mechanics involved in the droplet surface crater formation of the air-water interface of the impacting droplet on an immiscible high viscous liquid pool. The detailed characterization for various experimental parameter range is outside the scope of the present work.

\begin{figure}
  \centerline{\includegraphics[scale=1]{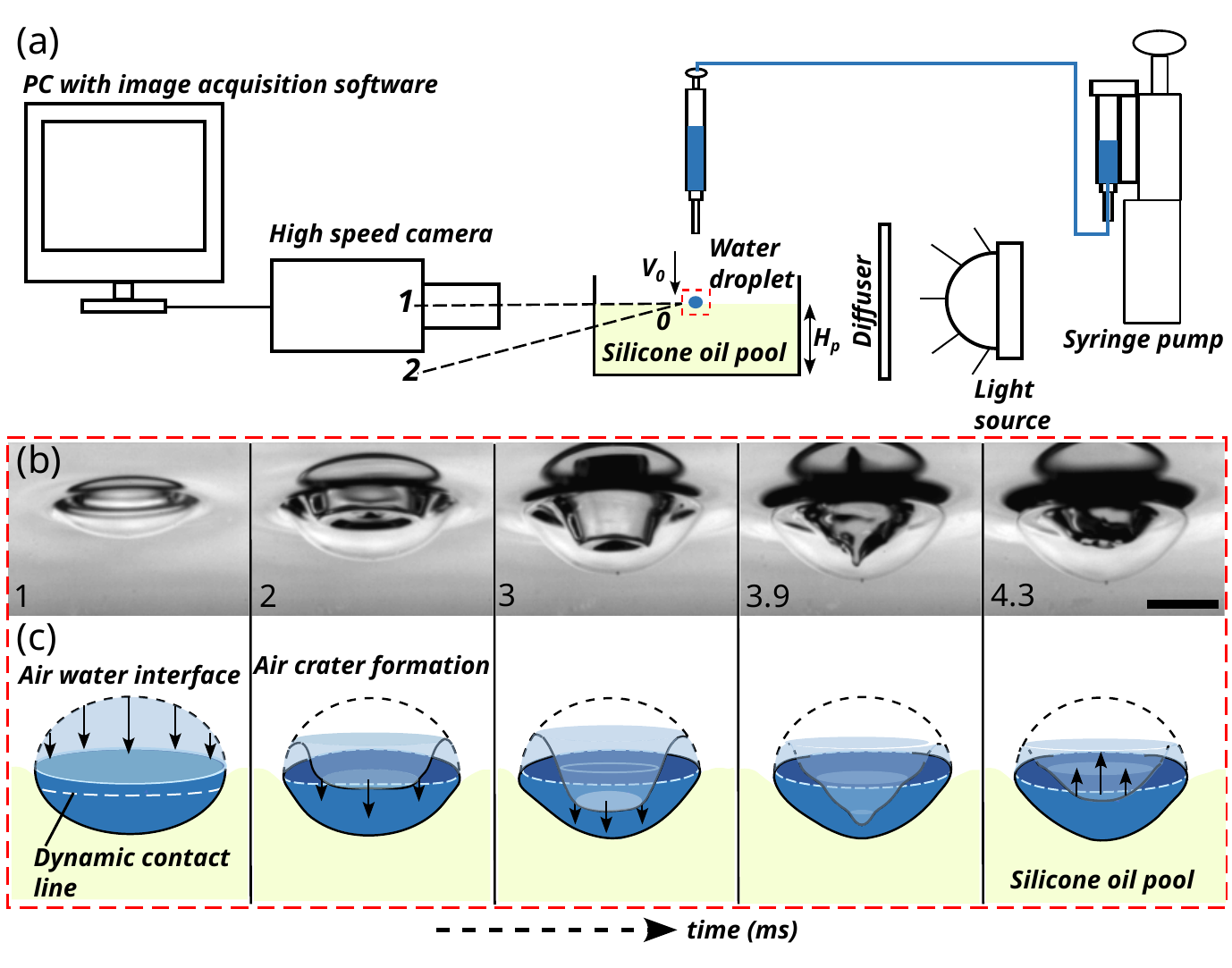}}
  \caption{
(a) Schematic of the experimental setup used for high-speed imaging (not to scale). The various components of the setup are labelled in the figure. The red dotted rectangle represents the region of interest. The dotted straight line $1-0$ and $2-0$ represents the visual axis (different sightline) of the camera used during high speed imaging experiments.
(b) Time sequence images for impact Weber number $We=65$ depicting the formation of air craters . Scale bar represents $1.56mm$ and timestamps are in milliseconds. 
(c) Schematic representing the air-water interface deformation process as a function of time leading to the formation of air craters. The arrows depict the direction of air-water interface propagation. The white line depicts the dynamic contact line that travels towards the north pole relative to the impacting droplet.}
\label{Figure1}
\end{figure}
 \section{Materials and Methods}
 Fig. 1(a) shows a schematic representation of the experimental setup
 for the high speed imaging. 
 The various components
 labelled in the  schematic representation are a syringe pump  (New Era Pump Systems, NE-1010); syringe and a hypodermic needle;  stroboscopic light source; a diffuser plate; water droplet; acrylic container; silicone-oil pool; computer with image acquisition software and high-speed cameras (Photron Fastcam Mini UX100, Photron Fastcam SA5) with Tokina and Navitar zoom lens. The small red dotted rectangle represent the region of interest for all experimental imaging. The dotted straight line $1-0$ and $2-0$ represents the visual axis of the high speed camera for image acquisition. 
 De-ionised water droplet of radius
 ($R_0=1.1{\pm}0.1mm$) 
 was used as the impacting liquid droplet. Silicone oil was used as the liquid pool, and the pool depth was maintained constant at $H_p{\sim}5mm$. The kinematic viscosity of the pool used was ${\nu}_{s}=350cSt$ and density was ${\rho}_s{\sim}970kg/m^3$. All experiments were conducted at room temperature of $T{\sim}298K$. The droplet impacts energy was varied by changing the impact velocity 
 ($0.3-2.2{\:}m/s$)
 , and hence 
 the triplet of non-dimensional numbers impact Weber number, Capillary number and Froude number ($We$, $Ca$, $Fr$) 
 was varied by changing the free fall height of the droplet from the free surface of the liquid pool using a vertical linear stage. We explored the 
 formation mechanism of air crater on the surface of the droplet
 in regime $(We,Ca,Fr)=(4-145,1.7-10.27,3.48-20.97)$.
 The high-speed imaging was performed at 
 $10kHz$ ($10^4$ frames per second)
  with a spatial resolution of $1.5$ micrometers per pixel (equivalent pixel density of ${\sim}{\:}666.667{\:}pixels/mm$). 
Fig. 1(b) shows a high-speed time sequence images depicting the formation of surface craters on the impacting droplet at impact Weber number $We=65$. The timestamps are in milliseconds and the scale bar represents $1.56mm$
The acquired high speed images were processed through an image processing pipeline consisting of background subtraction, image segmentation through an adaptive thresholding operation, followed by binarization and edge detection.
The geometrical and kinematic quantities like droplet size and velocity were computed through various imaging processing shape descriptors using Image processing
software
ImageJ \citep{schneider2012nih}
and, various in-house image processing codes written in Python \citep{van2009python} utilizing image processing libraries like opencv \citep{bradski2008learning,opencv_library,2015opencv} and scikit-image \citep{van2014scikit}.
The geometrical parameters 
extracted 
from the images that were taken from 
an oblique visual axis of the camera (Fig. 1(a)) 
were corrected using appropriate trigonometric and affine transformations \citep{che2018impact,chityala2020image}.
\section{Results and Discussions}
\subsection{Global Overview}
We discover that the mechanism behind air crater formation is (Fig. 1(c)) intricately related to the penetration dynamics of the impinging droplet through the liquid pool and depends on the impact Weber number. 
Fig. 1(c) depicts the schematic representation of the surface crater formation on the air-water interface of the impacting droplet. 
Fig. 2 shows the high-speed snapshots of a water droplet impacting a silicone liquid pool at various impact Weber numbers ($We=4,{\:}16,{\:}65,{\:}145$). 
The timestamps are in milliseconds, and the scale bars represent $1mm$.
The interaction between the silicone pool and water droplet occurs through the interplay of viscous, buoyancy, droplet weight, air-water, and water-oil surface tension forces. Beyond a critical impact Weber number ($We_c{\sim}10$), air craters of significant depths (crater depth comparable to droplet size) are formed. 
No significant air craters are observed during the penetration phase for impact Weber number smaller than the critical Weber number ($We<We_c$). The penetration time scale is largely increased due to the lubrication effect of the entrapped air layer formed under the droplet. The air layer rupture occurs at a time scale of the order of $t_r{\sim}2{\times}10^{-1}s$. The transient dynamics of the droplet (surface oscillation) occur early above the pool for $t<t_r$. For $t>t_r$, the droplet penetrates smoothly without any surface oscillation. The deceleration of the droplet is delayed due to the entrapped air and is non-impulsive. The retarding force during the penetration phase is of the order of $F{\sim}\mathcal{O}(4{\times10^{-4}})N$ for $We=4$.
The air craters formed for $We>We_c$ occur due to the rapid deceleration experienced by the impacting droplet over a short period (impulsive deceleration).
For impact Weber number larger than the critical Weber number ($We>We_c$), the air layer rupture time scale is of the order of $t_r{\sim}\mathcal{O}(5{\times}10^{-4})s$,  about two orders of magnitude faster than for low Weber number ($We{\sim}1$). We have shown using scaling analysis that the force responsible for the sudden deceleration is the viscous drag. The viscous drag force is of the order of $F_v{\sim}\mathcal{O}(1.5{\times}10^{-3}N)$ for $We=16$ which is comparatively larger than the buoyancy/weight of the droplet ($\mathcal{O}(5{\times}10^{-5}{\:}N)$). Further, we deduce a criterion for significant air crater depths. We show that the ratio of stokes viscous drag pressure to capillary pressure (${\Gamma}={\Delta}p_s/{\Delta}p_c$) is the parameter characterizing the formation and the size of air craters. ${\Delta}p_s/{\Delta}p_c{\sim}1$ corresponds to low Weber number states ($We{=4}$). The ratio ${\Gamma}$ increases as we increase the impact Weber number. The shape of the air crater central jet depends on the response of the droplet subjected to impulsive retardation. We show that the shape of the central air jet can be approximated with a fourth-order Legendre polynomial. We observe that the air crater formation/response time scales for all Weber number greater than the critical lies within the capillary time scale of $t_{c}{\sim}\sqrt{m/{\sigma}}{\sim}{\mathcal{O}(9{\times}10^{-3}s)}$. However, a complete air crater retraction time scale ($T$) is a monotonic function of impact Weber number ($T{\sim}We^{1/2}$). We further demonstrate that the immiscible impact problem could be mapped to
a constrained Rayleigh droplet model by incorporating a dynamic contact line relative to the impacting droplet. The crater retraction time scale increases with an increase in impact Weber number. When the air crater dynamics subsides, the droplet penetrates slowly under the influence of weight and buoyancy and attains an almost steady-state submerged configuration.

\begin{figure}
  \centerline{\includegraphics[scale=1]{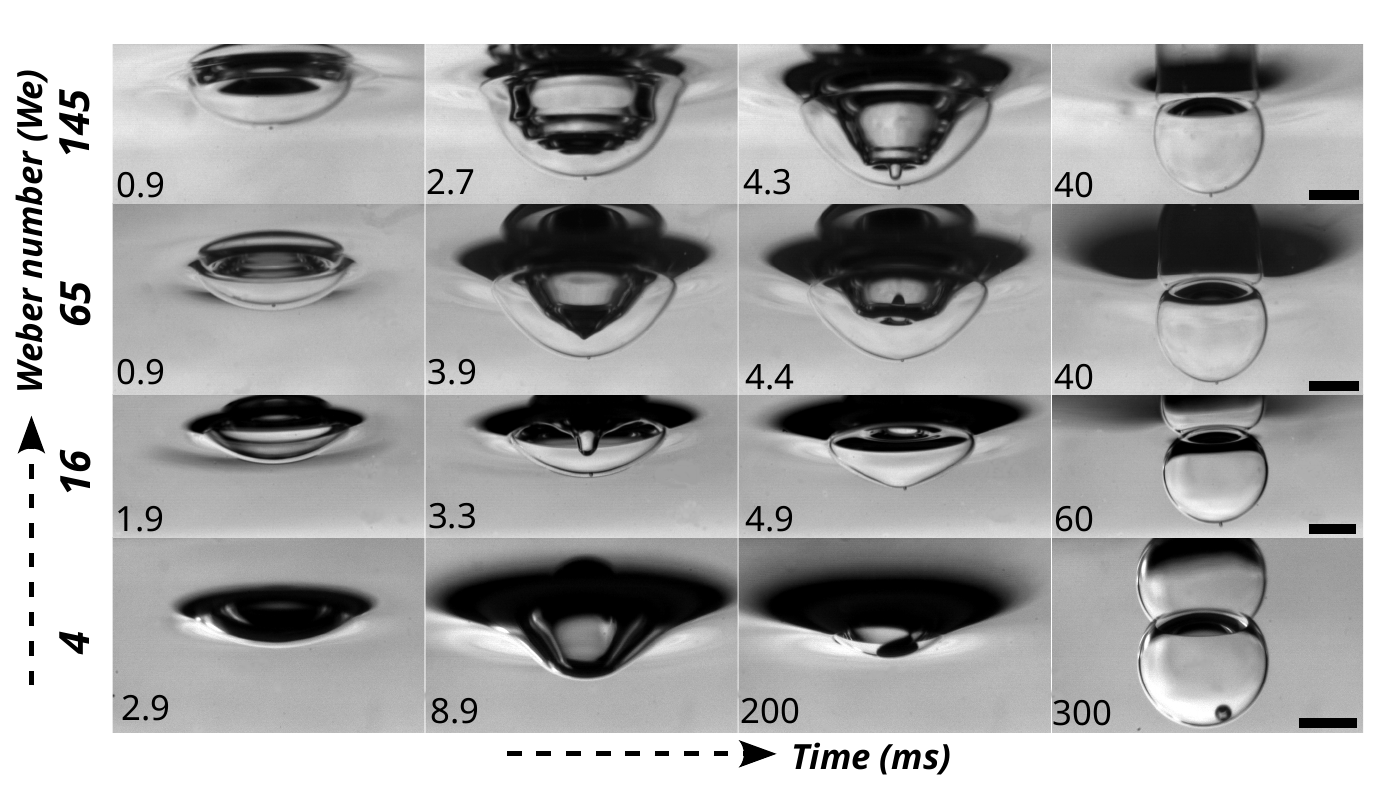}}
  \caption{Regime diagram showing the effect of impact Weber number on the evolution of droplet and air crater dynamics. The timestamp mentioned are in milliseconds and the scale bar denotes $1mm$.}
\label{Figure2}
\end{figure}

\subsection{Droplet penetration dynamics (Scales of various forces acting on the droplet)}
Fig. 3(a) depicts a schematic representation of an impacting droplet on a silicone oil pool. The oil pool is shown in light yellow. The schematic is drawn at a time instant $t$ ($t>t_r$) greater than the air rupture time scale, and the droplet is at a partially submerged state. $CM$ represents the centre of mass of the impinging droplet. 
The contact line of three different fluids (air, water, and oil) is marked as a white dotted line. The submerged portion/volume of the droplet immersed below the oil surface is quantitatively characterized and measured by the penetration depth $h(t)$ and penetration width $w(t)$ (Fig. 3(a)).
We assume that the droplet deforms into an ellipsoid during the penetration process. The ellipsoid is characterized by the semi-major axis $a(t)$ and the semi-minor axis $b(t)$ respectively (Fig. 3(a)). The forces acting on the droplet are the weight $F_{mg}$ downwards, buoyancy force $F_b$ upwards, and the viscous drag force $F_v$ upwards. The coordinate axis $z$ is chosen to be positive downwards. 
From mass conservation of the impacting droplet we have
\begin{equation}
    {\rho}_w\frac{4}{3}{\pi}R_0^3={\rho}_w\frac{4}{3}{\pi}a^2(t)b(t)=m
\end{equation}
where ${\rho}_w$ is the density of the impacting droplet (water here), $R_0$ is the initial radius of the droplet and $m$ is the mass of the impacting droplet. For droplet radius of $R_0=1.1mm$, the mass of the impinging water droplet is $m{\sim}\mathcal{O}({5.58{\times}10^{-6}}kg)$.
Simplifying equation (3.1) we have
\begin{equation}
    a^2(t)b(t)=R_0^3
\end{equation}
The dynamical equation (Newton's second law of motion) for the impinging liquid droplet is given by 
\begin{equation}
F_{mg} - F_b(t) - F_v(t) = m\frac{dV_{CM}(t)}{dt} = ma_{CM}
\end{equation}
with initial condition for $V_{CM}$ as
\begin{equation}
    V_{CM}(t=0)=V_0
\end{equation}
where $V_{CM}$ is the centre of mass velocity, $V_0$ is the impact velocity. The $t=0$ defines the time instant just before the penetration of the droplet.
$F_{mg}=mg$ is the weight of the impacting droplet where $g=9.81{\:}m/s^2$ is the acceleration due to gravity, $F_b$ is the buoyant force on the droplet and $F_v$ is the viscous drag force on the droplet due to the interaction with the silicone oil pool.
For $R_0{=}1.1mm$, the weight of the droplet is of the order of $F_{mg}{\sim}\mathcal{O}(5.47{\times}10^{-5}{\:}N)$. The submerged droplet volume in the silicone oil pool can be approximated by an ellipsoidal cap approximation (refer to Fig. 3(a))
\begin{equation}
    v(t) = \frac{{\pi}a^2(t)h^2(t)}{3b^2(t)}(3b(t)-h(t))
\end{equation}
where $h(t)$ is the depth of penetration of the droplet.
\begin{figure}
  \centerline{\includegraphics[scale=1]{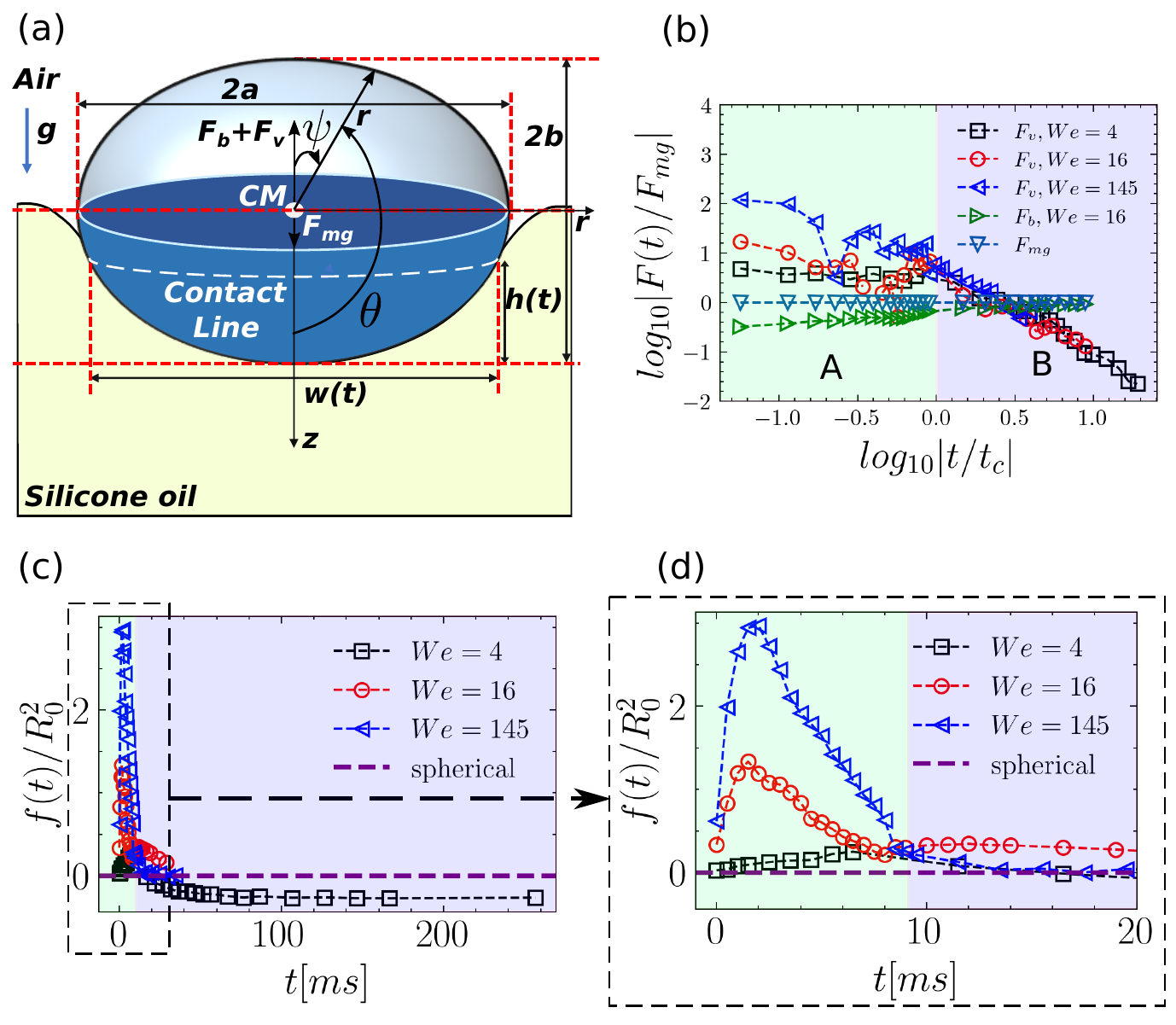}}
  \caption{(a) Schematic representation depicting various geometrical scales and forces acting on the impacting droplet while it penetrates the silicone oil pool. (b) log-log plot depicting the strength of various forces in comparison to weight of the droplet as a function of time normalized with respect to capillary time scale $t_c{=}\sqrt{m/{\sigma}_{aw}}$, where $m$ is the mass of the droplet and ${\sigma}_{aw}$ is the air-water surface tension with impact Weber number as a parameter. (c) and (d) $f(t)/R_0^2$ as a function of time in millisecond. The function $f(t)$ is a measure of a spherical drop shape. }
\label{Figure3}
\end{figure}
The buoyancy force is the weight of the displaced silicone oil, and therefore $F_b(t)$ becomes
\begin{equation}
    F_b(t)=v(t){\rho}_sg=\frac{{\pi}a^2h^2{\rho}_sg}{3b^2}(3b-h)
\end{equation}
Using equation (3.2) in (3.6) the buoyancy force becomes
\begin{equation}
    F_b(t) = \frac{{\rho}_sg{\pi}R_0^3h^2}{3b^3}(3b-h)
\end{equation}
 At $h=b$ (i.e. when the droplet (ellipsoid) is half submerged) the buoyancy force becomes
\begin{equation}
    F_b = \frac{2}{3}{\rho}_s{\pi}R_0^3g{\sim}\mathcal{O}(2.65{\times}10^{-5}{\:}N)
\end{equation}
Notice that the order of magnitude of buoyancy force is the same as the weight of the droplet, and the buoyancy force becomes important at long time scales to determine the maximum steady-state submerged depth of the droplet.
During the early stage of impact ($t{\sim}2ms$) the semi-major axis is approximately equal to the semi-minor axis ($a{\simeq}b$) and hence the order of magnitudes of $a$ and $b$ are $\mathcal{O}(R_0)$. Equation (3.6) can therefore be approximated for computing the buoyancy force at early impact time as
\begin{equation}
    F_b(t)=v(t){\rho}_sg{\sim}\frac{{\pi}h^2{\rho}_sg}{3}(3R_0-h)
\end{equation}
\begin{figure}
  \centerline{\includegraphics[scale=1]{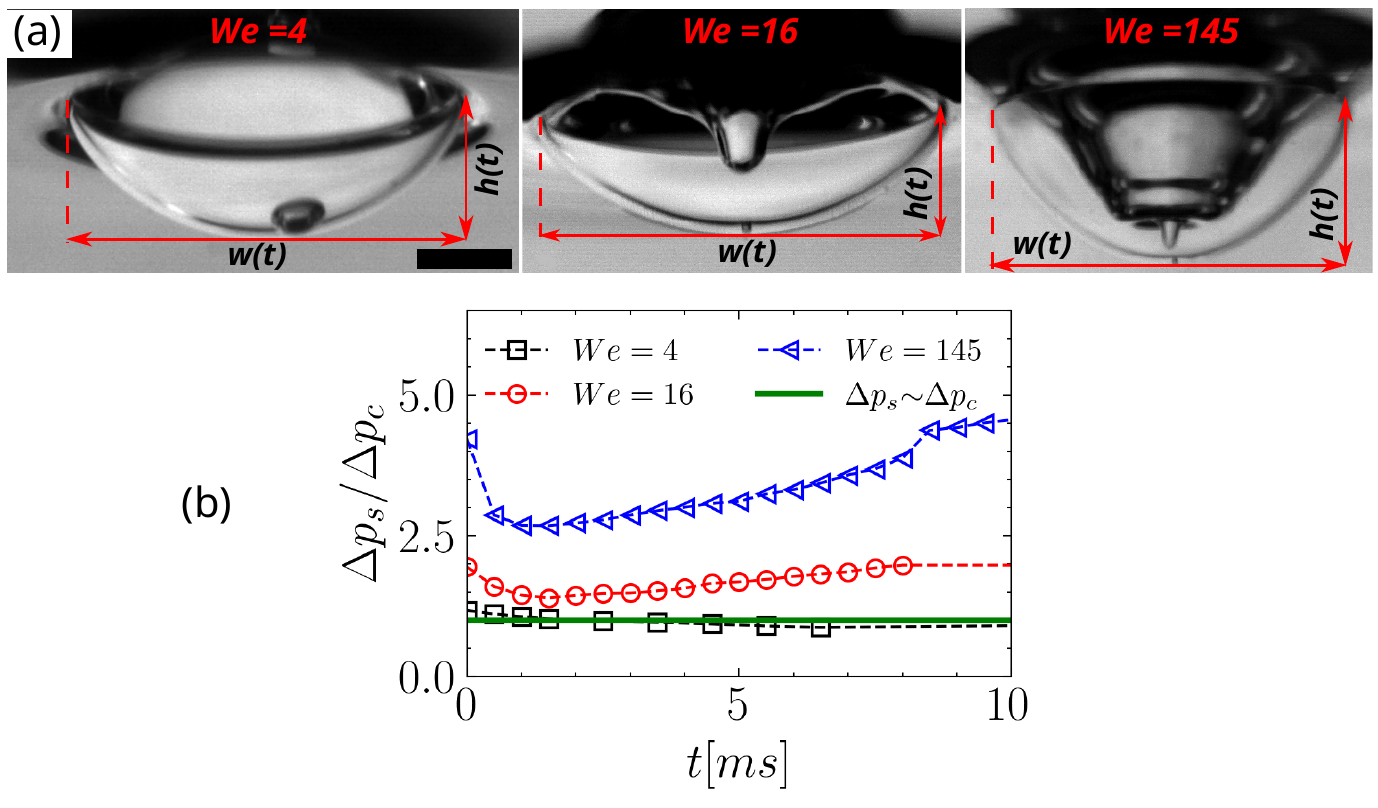}}
  \caption{(a) Snapshots depicting the penetration depth ($h(t)$) and width ($w(t)$) respectively for impact Weber number of $4$, $16$ and $145$.
  (b) ${\Gamma}(t)={{\Delta}p_s}/{{\Delta}p_c}$ as a function of time ($t_r<t<t_c$) for impact Weber number of $4$, $16$ and $145$ respectively.}
\label{Figure4}
\end{figure}
The viscous force on the droplet $F_v$ can be approximated by the Stoke's drag law \citep{moghisi1981experimental} as given by
\begin{equation}
    F_v(t){\sim}3{\pi}{\mu}_sw(t)\frac{dh}{dt}
\end{equation}
where ${\mu}_s$ is the viscosity of the silicone oil pool. Fig. 3(b) represents the scales of viscous force, buoyancy forces comapred to the weight of the droplet plotted in a logarithmic scale against logarithmic time normalized with respect to the capillary time scale $t_c$ for impact Weber number of $4$, $16$ and $145$.
Stoke's approximation used in equation (3.10) becomes valid since the Reynolds number inside the silicone oil pool is very small ($Re=V_0R_0/{\nu}_s{\sim}1$). The maximum viscous force for $We{\sim}16$  scales as $F_v{\sim}\mathcal{O}(1.5{\times}10^{-3}{\:}N)$ in the initial phase of penetration. Note that the viscous drag initially is a hundred times larger than the buoyant force and weight of the droplet. Therefore the initial dynamics of droplet penetration and droplet response are governed by Stoke's viscous drag. 
\subsection{Characterizing droplet deformation from spherical shape}
In general as the droplet penetrates the liquid pool, the droplet undergoes deformation and evolves as an ellipsoid (Fig. 3(a)). We have measured and characterized the amount of deviation from a spherical shape by calculating the function $f(t)$ defined as
\begin{equation}
    f(t) = h^2(t) + \frac{w^2(t)}{4} - 2R_0h(t)
\end{equation}
$f(t)=0$ represents a spherical undeformed droplet while the penetration occurs. Larger non zero values of $f(t)$ depicts the deviation from spherical symmetry. $f(t)$ has a dimension of square length and therefore can be represented in a dimensionless way by normalizing with respect to $R_0^2$. Equation (3.11) therefore becomes
\begin{equation}
    \frac{f(t)}{R_0^2}=\frac{h^2(t)}{R_0^2}+\frac{w^2(t)}{4R_0^2} - 2\frac{h(t)}{R_0}
\end{equation}

Fig. 3(c) and 3(d) shows the evolution of $f(t)/R_0^2$ plotted as a function of time using equation (3.12) for various impact Weber number ($We=4,16,145$). It can be inferred from the variation of $f(t)$ that for low Weber numbers ($We{\sim}4$), the droplet shape while penetration is very close to spherical. However, increasing the impact Weber number ($We{\sim}16,145$) the droplet deformation is substantial as can be observed by the non-zero values of $f(t)/R_0^2>1$. 
\begin{figure}
  \centerline{\includegraphics[scale=1]{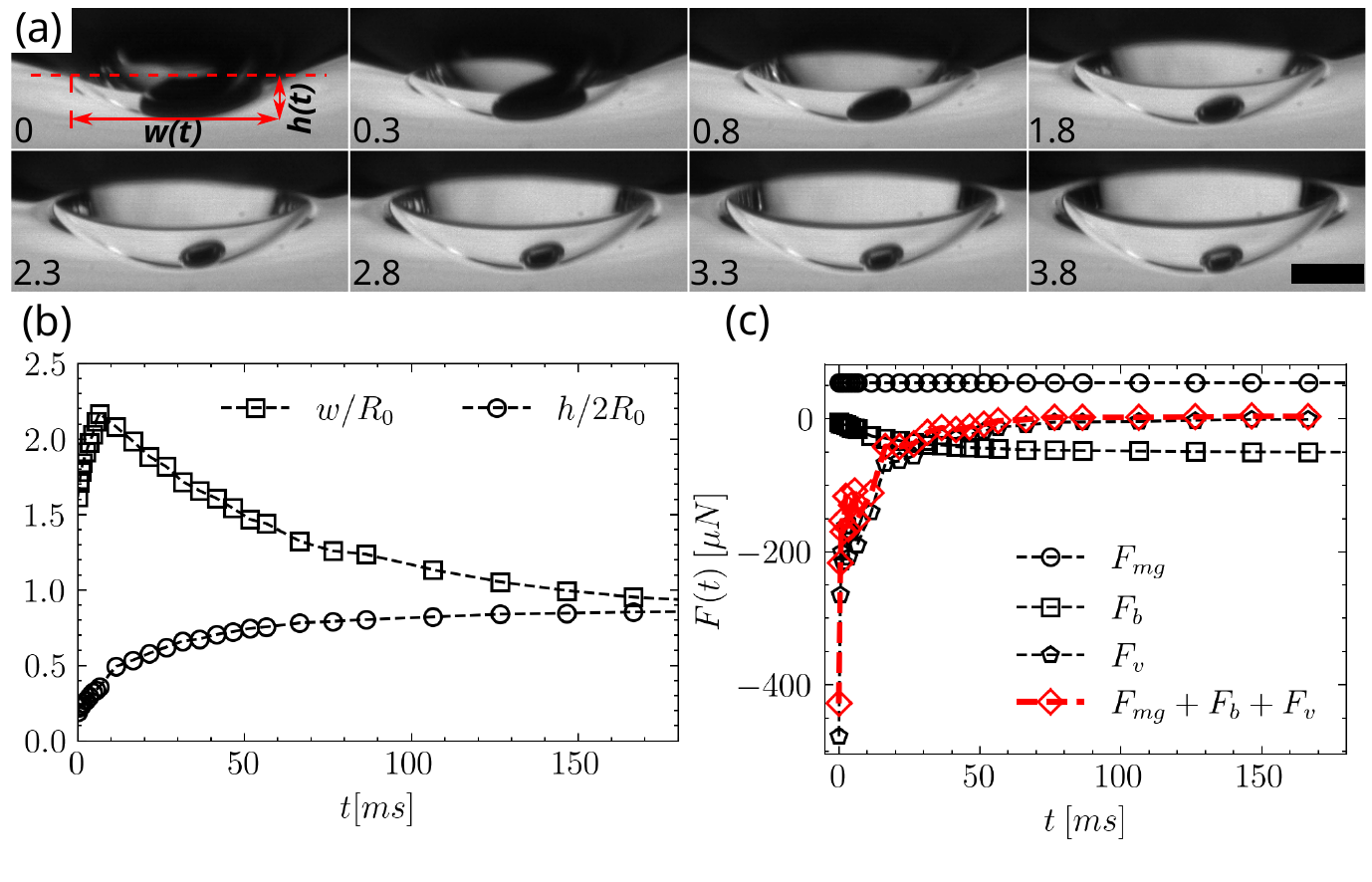}}
  \caption{(a) High speed snapshots depicting the evolution of air layer rupture forming a bubble during penetration through silicone oil for impact Weber number $We=4$. The timestamps are in milliseconds and the scale bar represents $500{\mu}m$. (b) Temporal evolution of kinematic parameters like penetration width and depth. (c) Temporal evolution of the various forces acting on the impacting droplet. The resultant of the various forces is also shown in red. \href{https://drive.google.com/file/d/1br86bA7VTtk_IbnuqM-jIjIYKRa7HbjR/view?usp=sharing}{Refer to supplementary Movie 1 for $We=4$ here.}}
\label{Figure5}
\end{figure}
Regardless of the impact Weber number, beyond the capillary time $t_c$, the droplet penetrates, maintaining a spherical shape. It could be inferred that the deviation from the spherical shape due to penetration is a short time scale phenomenon ($t<t_c$).  
The dominant force experienced by the droplet on penetration is the viscous force (refer to Fig. 3(b)). We observe the formation of significant air craters due to sudden viscous drag acting on the droplet beyond critical impact Weber number of $10$. 
Fig. 4(a) depicts the geometrical characteristics of the deforming droplet/air craters formed during penetration for three different impact Weber numbers equal to $4$, $16$, and $145$. We observe that significant air craters does not form for low Weber number ($We{\sim}4$ shown here) due to delayed air layer rupture ($t_r{\sim}\mathcal{O}(2{\times}10^{-1}s)$). The impulsive nature of the drag force does not manifest during the penetration process due to delayed air layer rupture, and the viscous force becomes gradual for $We<We_c$.
Air craters of significant shape and size are observed for larger Weber numbers ($We{\sim}16$ and $We{\sim}145$ shown here).
\subsection{Air crater formation criterion}
The air craters observed for $We>We_c$ (refer to Fig. 4(a)) occur due to impulsive viscous drag force acting on the droplet across the immersed surface/volume in the silicone oil bounded by the contact line (refer to the schematic shown in Fig. 3(a). 
We introduce a non-dimensional parameter ${\Gamma}(t)$ as the ratio of the stokes drag pressure to capillary pressure across the air water interface in the air crater formation time scale ($t_r<t<t_c$). The parameter can be represented as 
\begin{equation}
    {\Gamma}(t)=\frac{{\Delta}p_s}{{\Delta}p_c} {\sim} \frac{{\mu}_sw(t)V_0R_0}{{\sigma}_{aw}w^2(t)}
\end{equation}
where ${\Delta}p_s{\sim}{\mu}_sw(t)V_0/w^2(t)$ is the pressure force due to the stokes drag and ${\Delta}p_c{\sim}{\sigma}_{aw}/R_0$ is the capillary pressure across the air-water interface.
${\Gamma}(t){\sim}1$ signifies that the impulsive drag is almost balanced out by the capillary force acting across the air-water interface of the droplet. 
\begin{figure}
  \centerline{\includegraphics[scale=1]{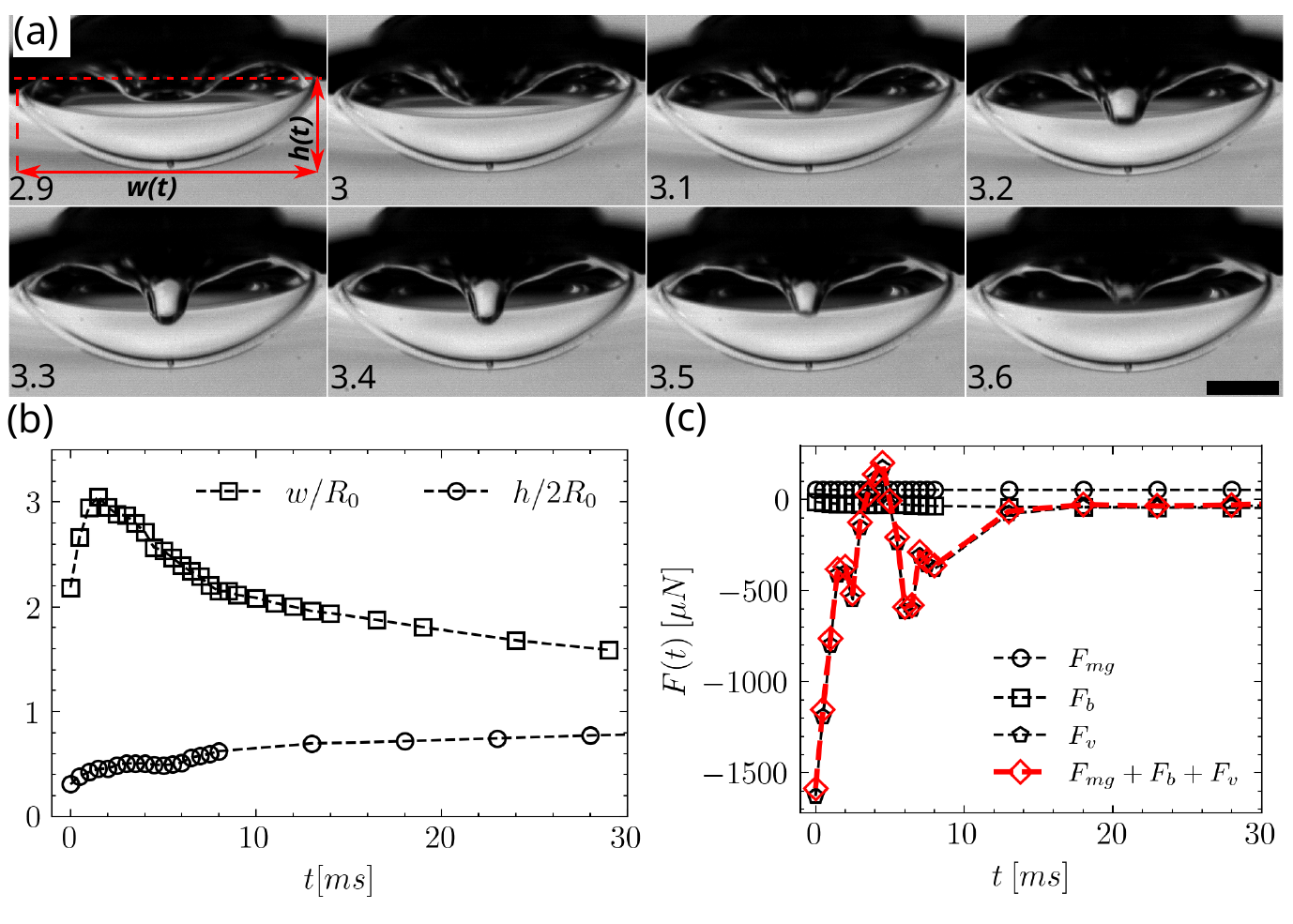}}
  \caption{(a) High speed snapshots depicting the evolution of air layer rupture forming a bubble during penetration through silicone oil for impact Weber number $We=16$. The timestamps are in milliseconds and the scale bar represents $500{\mu}m$. (b) Temporal evolution of kinematic parameters like penetration width and depth. (c) Temporal evolution of the various forces acting on the impacting droplet. The resultant of the various forces is also shown in red. \href{https://drive.google.com/file/d/1D-0nSXErJFs-BQyQ-JfryQzGSbH8ZAQ9/view?usp=sharing}{Refer to supplementary Movie 2 for $We=16$ here.}}
\label{Figure6}
\end{figure}

Detectable air craters during the penetration process can only be observed if the viscous force becomes comparable and overcomes the capillary force across the droplet interface, which corresponds to ${\Gamma}(t)>1$.
We also show that for ${\Gamma}(t){\leq}1$ for $t<t_c$, no significant air craters are observed, as is the case for low impact Weber number. Increasing Weber number increases ${\Gamma}(t)$ and hence the ratio of Stokes viscous force to capillary force across the air-water interface of the droplet. Equation (3.13) can be rewritten in terms of the capillary number $Ca={\mu}_sV_0/{\sigma}_{aw}$ as

\begin{equation}
    {\Gamma}(t)=\frac{{\Delta}p_s}{{\Delta}p_c} {\sim} Ca\frac{R_0}{w(t)}
\end{equation}

Figure 4(b) represents ${\Gamma}(t)={{\Delta}p_s}/{{\Delta}p_c}$ as a function of time for $t_r<t<t_c$ for Weber numbers $4$, $16$ and $145$. It can be inferred from Fig. 4(b) that significant air craters are formed for ${\Gamma}>1$ where the impulsive viscous forces become larger than the capillary force across the air-water interface. For small Weber number, $We<We_c=10$, ${\Gamma}{\lesssim}1$ signifying that the viscous forces are smaller than the capillary force across the air-water interface due to surface tension.
\subsection{Evaluating the centre of mass velocity ($V_{CM}$)}
We observe that the order of magnitude of air crater jet velocity is comparable to the centre of mass velocity of the droplet ($V_{CM}$). 
Using equation (3.9) and (3.10) for the buoyancy and viscous terms respectively in equation (3.3), the centre of mass acceleration of the droplet can be expressed as
\begin{equation}
    \frac{dV_{CM}(t)}{dt}{=}\frac{1}{m}\left(mg - \frac{{\pi}h^2{\rho}_sg}{3}(3R_0-h) - 3{\pi}{\mu}_sw(t)\frac{dh}{dt}\right)
\end{equation}

Integrating equation (3.15) we have
\begin{equation}
    \int_{V_0}^{V_{CM}(t)}{\frac{dV_{CM}(t)}{dt}}dt= \int_{0}^{t}\frac{1}{m}\left(mg - \frac{{\pi}h^2{\rho}_sg}{3}(3R_0-h) - 3{\pi}{\mu}_sw(t)\frac{dh}{dt}\right)dt
\end{equation}
Simplifying equation (3.16) we have
\begin{equation}
    V_{CM}(t) = V_0 + gt -  \frac{{\pi}{\rho}_sg}{3}\int_0^th^2(3R_0-h)dt - 3{\pi}{\mu}_s\int_0^tw(t)\frac{dh}{dt}dt
\end{equation}
Notice that the acceleration due to gravity term $gt$ and the buoyancy term (third term on the right hand side) is negligible compared to the viscous term (last term on the right hand side) for $t_r<t<t_c$. The last term representing the viscous forces is the most dominant for $t_r<t<t_c$.

\begin{figure}
  \centerline{\includegraphics[scale=1]{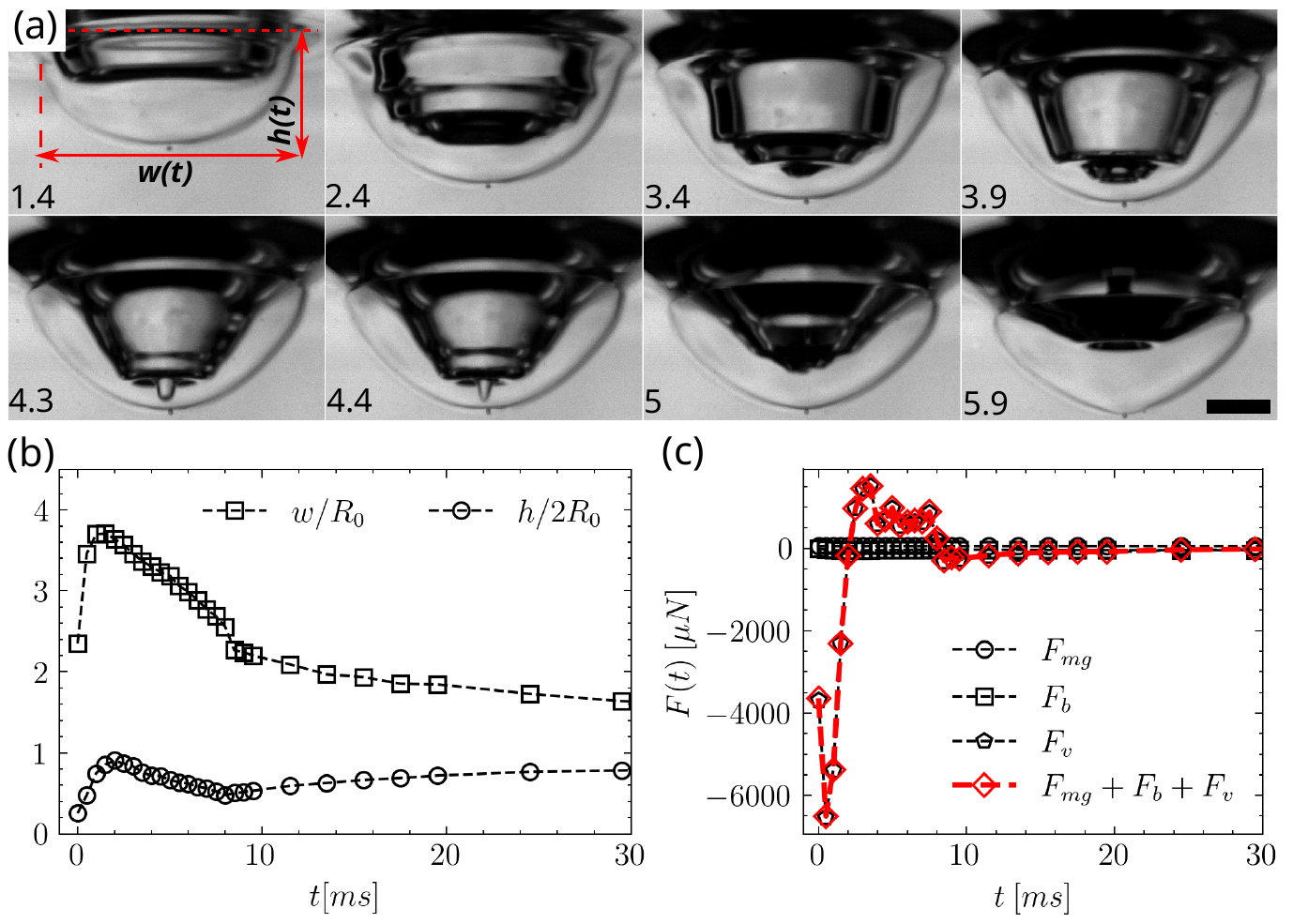}}
  \caption{(a) High speed snapshots depicting the evolution of air layer rupture forming a bubble during penetration through silicone oil for impact Weber number $We=145$. The timestamps are in milliseconds and the scale bar represents $500{\mu}m$. (b) Temporal evolution of kinematic parameters like penetration width and depth. (c) Temporal evolution of the various forces acting on the impacting droplet. The resultant of the various forces is also shown in red. \href{https://drive.google.com/file/d/1PFcpJvw--YI1FTmh2AEwRRRa8DEz_Bly/view?usp=sharing}{Refer to supplementary Movie 3 for $We=145$ here.}}
\label{Figure7}
\end{figure}

\begin{figure}
  \centerline{\includegraphics[scale=1]{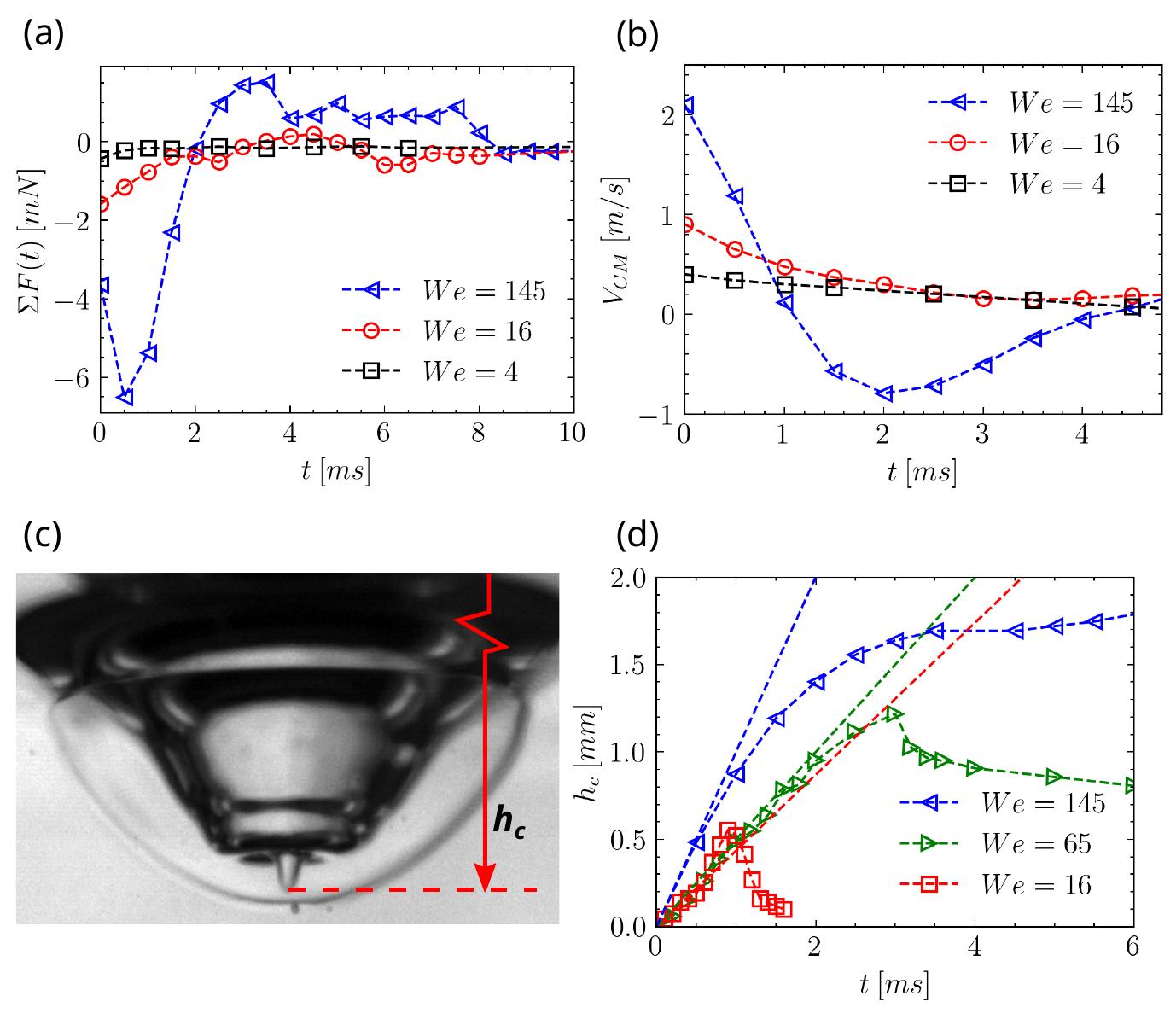}}
  \caption{(a) Comparison of the resultant force ${\Sigma}F(t)$ in ${mN}$ acting on the droplet during the penetration process as a function of time $t$ in ${ms}$ for impact Weber number ($We$) of $4$, $16$ and $145$ evaluated using the right hand side of equation (3.15). (b) Comparison of the centre of mass velocity of the droplet ($V_{CM}$) in ${m/s}$ plotted as a function of time $t$ in ${ms}$ for impact Weber number ($We$) of $4$, $16$ and $145$. (c) A high speed snapshot depicting the crater depth $h_c$. (d) Crater depth ($h_c$) evolution as a function of time in $ms$.}
\label{Figure8}
\end{figure}

\begin{figure}
  \centerline{\includegraphics[scale=1]{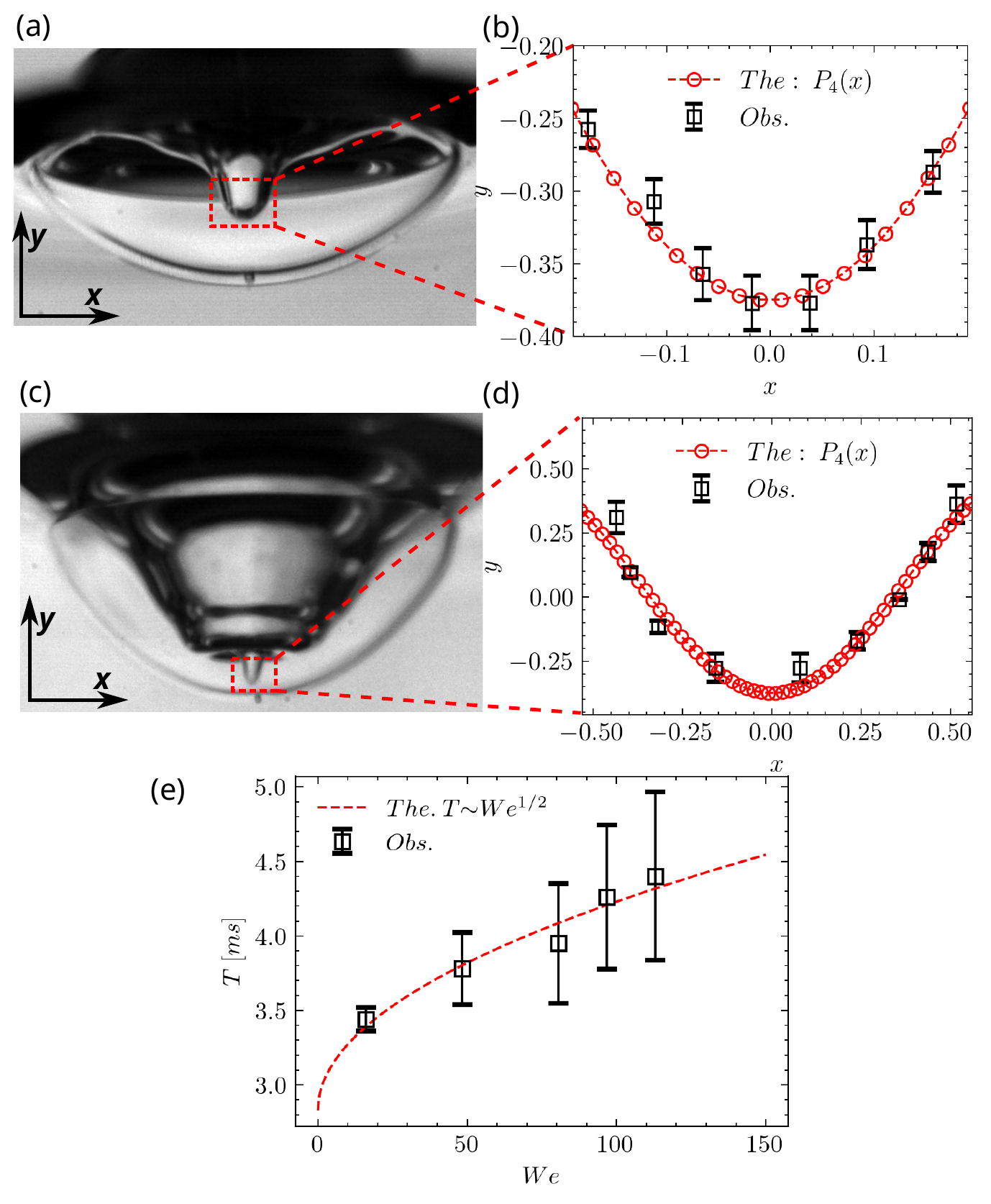}}
  \caption{(a) Image depicting the profile of the reversed jet formed during air crater formation/collapse at impact Weber number of 16. (b) Comparison of the experimental profile of the central jet with 4-th degree Legendre polynomial.
  (c) Image depicting the profile of the reversed jet formed during air crater formation/collapse at impact Weber number of 145. (d) Comparison of the experimental profile of the central jet with 4-th degree Legendre polynomial. (e) Experimental and theoretical comparison of air crater time scale ($T$) as a function of impact Weber number ($We$) based on equation (3.22).}
\label{Figure9}
\end{figure}
Fig. 5, Fig. 6, and Fig. 7 show various kinematic and dynamic quantities for impact Weber numbers of $4$, $16$, and $145$, respectively. Fig. 5(a) depicts the high-speed snapshots during the penetration process for an impact Weber number of $4$. The penetration dynamics for low Weber numbers are drastically different as the droplet, and air layer interaction is significant. The air layer rupture and formation of a subsequent bubble can be observed in Fig. 5(a). Further, as the droplet penetrates through the silicone oil pool, the droplet shape is close to spherical (refer to Fig. 3(c) and Fig. 3(d)). Fig. 5(b) shows the temporal evolution of submerged droplet normalized width ($w(t)/R_0$) and normalized depth ($h(t)/2R_0$) respectively. Using the experimental value of $w(t)$ and $h(t)$, we have calculated the resultant force acting on the droplet. All the individual forces ($F_{mg}$, $F_b$ and $F_v$) along with the resultant force (in red $F_{mg}+F_b+F_v$) is plotted in Fig. 5(c). Notice that during the early time of penetration for $t<t_c$, the dominant force is only the viscous drag. In general, we observe that the resultant force is almost entirely due to viscous drag with buoyancy and weight almost balancing each other as time increases. Using the resultant external forces acting on the droplet (Fig. 5(c)), the centre of mass velocity ($V_{CM}$) can be evaluated using equation (3.17). We evaluate centre of mass velocity ($V_{CM}$) for early penetration time ($t<t_c$) as the approximation under which equation (3.17) is derived becomes increasingly accurate for small time duration. The effect of the impact Weber number can be understood by analyzing the scale of $F_v$ as a function of the Weber number. It can be inferred from Fig. 3(b) that on increasing the impact Weber number, the viscous force $F_v$ changes significantly. The resultant force on the droplet is approximately equal to that of the viscous force. Therefore on increasing the impact Weber number, the ratio of stokes viscous pressure to capillary pressure of the droplet ${\Gamma}$ increases resulting in the formation of air crater/reversed jets of various shapes and sizes (refer to Fig. 4(a) and Fig. 4(b)). Fig. 6(a) shows a sequence of high-speed snapshots showing the evolution of the air crater for $We=16$. Fig. 6(b) plots the evolution of normalized penetration depth and width of the impacting droplet for impact Weber number of $16$. The individual and resultant forces acting on the droplet are shown in Fig. 6(c). Fig. 7(a) shows the air crater formation and evolution for impact Weber number of $145$. The crater formed is significantly different compared to the low Weber number impact. For Weber number of $16$, a single air jet is formed due to the capillary waves interfering. Whereas for Weber number of $145$, a cascade of capillary waves is formed, forming a bigger intricate air crater. The air crater normalized depth ($h(t)/2R_0$) and normalized width ($w(t)/R_0$) is plotted as a function of time in Fig. 7(b). Using the experimental air crater geometrical parameters, the individual and the resultant force acting on the impacting droplet were calculated and are shown in Fig. 7(c). 
The resultant force ${\Sigma}F(t)$ acting on the impinging droplet during the penetration process is plotted as a function of time for various impact Weber number in Fig. 8(a) for comparison purposes ($We=4,16,145$). The centre of mass velocity ($V_{CM}$), the solution for equation (3.17) is plotted in Fig. 8(b) for various impact Weber numbers. The centre of mass velocity decreases as the droplet experiences retarding forces as it penetrates the liquid silicone oil pool.
We can further observe that the centre of mass velocity shows a damped oscillatory response for $t{\sim}t_c$.
Fig. 8(c) shows a typical air crater geometry and its characteristics scale (air crater depth $h_c$) formed on the surface of an impacting droplet. The evolution of air crater depth $h_c$ as a function of time $t$ is plotted in Fig. 8(d) for impact Weber numbers of $16$, $65$ and $145$.
Note that the initial slope of $h_c$ vs. $t$ plot is ordered according to the centre of mass velocity for various Weber numbers, respectively. $We=145$ has the highest initial slope followed by $We=65$ and $We=16$.
\subsection{Transient Droplet Response (Local Analysis)}
The air crater/jets are formed due to the transient response of the droplet to impulsive forces acting on the droplet. The sudden deceleration of the droplet induces droplet oscillation that causes air craters of various shapes and sizes to form. Further, relative to the droplet the external force acting could also be formulated based on the relative effect of the surrounding air field.
The transient response of the droplet surface based on sudden retardation can be characterized by surface displacement ${\eta}=r({\mu},t)/R_0$; where $r({\mu},t)$ is the radial location of the droplet interface at any angular coordinate ${\theta}$ at time $t$ (refer to Fig. 3(a)) and $R_0$ is the initial droplet radius. We closely follow this part of the analysis based on the work of Harper et al. and Simpkins et al. \citep{harper1972breakup,simpkins1972water}.
The perturbation result for the surface displacement based on a linear theory proposed by Harper et al.\citep{harper1972breakup} is given as
\begin{equation}
    {\eta} = 1 + {\epsilon}\sum_{n=0}^{\infty}\frac{n(2n+1)}{4{\beta}_n^2}C_nP_n({\mu})[cos({\beta}_nt)-1]
\end{equation}
where ${\epsilon}={\rho}_a/{\rho}_w$ is the density ratio of air to water characterizing the amplitude of the perturbation, $n$ is a positive integer and $C_n$ are the weighting coefficients given by,
\begin{equation}
    C_n = \int{P}_e^{(0)}P_n({\chi})d{\chi}
\end{equation}
where ${\chi}=cos{\psi}$.
$P_n(\mu)$ represents the n-th order legendre polynomials where ${\mu}=cos{\theta}$ and 
\begin{equation}
    {\beta}_n={\left(\frac{(n-1)n(n+2)}{We_{R_0}}\right)^{1/2}}
\end{equation}
where ${\beta}_n$ denotes the frequency of n-th Legendre mode and $We_{R_0}$ denotes the Weber number defined with respect to the initial radius $R_0$ ($We = 2 We_{R_0}$). $P_e^{(0)}$ represents the leading term of the external pressure $P_e$ in the perturbation expansion with ${\epsilon}$ as the perturbation parameter acting on the droplet during the penetration process.
\begin{equation}
    {\beta}{\sim}We^{-1/2}
\end{equation}
Refer to Fig. 3(a) for the definition of various angular coordinates like ${\theta}$ and ${\psi}$. Note that near the droplet top surface that corresponds to ${\psi}=0$ and ${\theta}={\pi}$, the droplet shape approximates that of a Legendre polynomial locally. Fig. 9(a) and 9(b) show the comparison of the air jet profile at the centre with the fourth-order Legendre polynomial for $We=16$. Similar comparison is shown in Fig. 9(c) and 9(d) for $We=145$. In principle, the combined air crater deformation can be thought of as the superposition of various Legendre modes. The period of half oscillation of the air jets scales as
\begin{equation}
    T{\sim}{\beta}^{-1}{\sim}We^{1/2}
\end{equation}

Therefore, from equation (3.22), we observe that the period of air crater dynamics increases monotonically with the impact Weber number. As inferred from Fig. 9(e), we observe the corresponding monotonic behavior experimentally.

\begin{figure}
  \centerline{\includegraphics[scale=1]{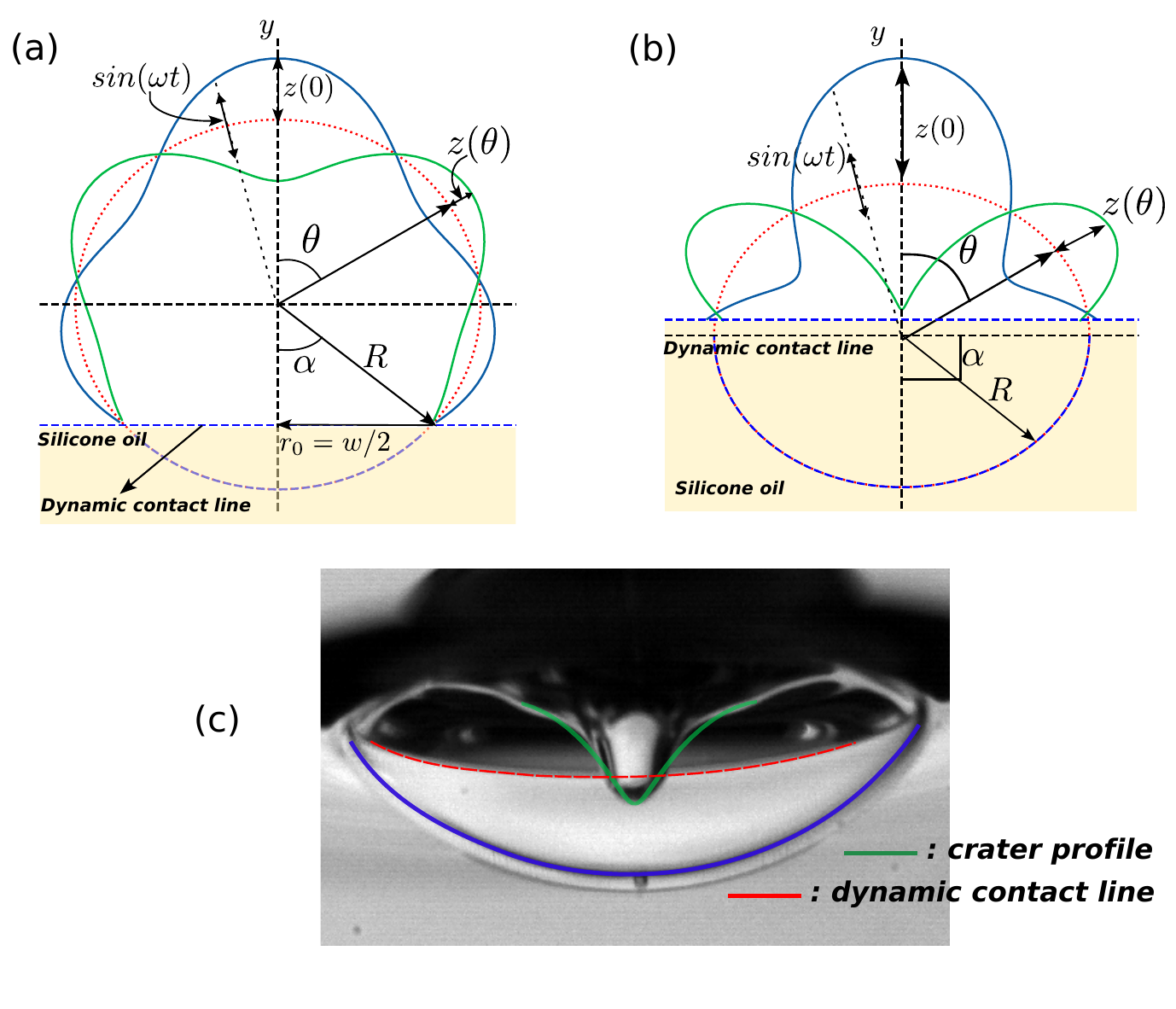}}
  \caption{Graphical representation depicting the geometrical nomenclature and the air-water inerfacial waves for the global deformation analysis. (a) for ${\alpha}={\pi}/4$ (b) for ${\alpha}={\pi}/2$}. (c) An experimental snapshot depicting the observed crater characteristics for impact Weber number $We=16$.
\label{Figure10}
\end{figure}

\subsection{{Transient vibrational response of the droplet based on constrained rayleigh drop model
(Global Analysis)}}
In this section we generalize the local analysis performed in the previous section (3.6) with a global response model. When the entrapped air beneath the impacting droplet ruptures and forms a central air bubble, a three-fluid dynamic contact line is formed (refer to Fig. 1(c), 3(a), 10(a) and 10(b)). The contact line is a one dimensional boundary separating the three immiscible fluids (air, water and silicone oil) and is one of the important aspects controlling the response of the droplet. The dynamic contact line separates the droplet interface into two disjoint curved surface areas. The dynamic contact line is characterized using an inclination angle ${\alpha}$ measured from the $-y$ axis.   
Fig. 10(a), 10(b) graphically represents the geometry of a penetrating droplet from constrained Rayleigh drop perspective.
Fig. 10(a) represents the droplet penetration at any given instant for an acute value of ${\alpha}={\pi/4}$. For ${\alpha}={\pi/2}$, the schematic is shown in figure 10(b). The colored saffron region represents the silicone oil pool. $R=R_0$ represents the radius of the undeformed equilibrium shape of the droplet. The dynamic contact line is represented by the horizontal dotted blue line. The curved dotted blue line represents the bottom submerged portion of the droplet.
The red dotted curved interface represents the equilibrium position of the air-water interface. The blue and green solid line depicts the oscillating air-water interface of the droplet that leads to the formation of air craters. $z({\theta})$ represents the amplitude of the air-water interface from the equilibrium position of the interface at a given polar angle ${\theta}$ and time $t$.
In general the dynamic contact line imposes the constraint of zero amplitude along its perimeter. However, owing to large viscosity of the silicone pool, the entire curved surface of the droplet submerged in the silicone pool (water-oil interface) could be considered as an surface area of negligible (zero) amplitude. The amplitude response of the droplet from the mean equilibrium shape in general depends on the impact Weber number and increases monotonically with the same (refer Fig. 8(d)). For small impact Weber number the droplet interface amplitudes and the deviation from the average spherical equilibrium shape is relatively small (refer Fig. 3(d)).  We therefore use the constrained rayleigh drop model for such low energy impacts. We follow the analysis of Strani, Bostwick-Steen closely in this section \citep{strani1984free,bostwick2009capillary}. In a recent work, Kern et al. showed that in drop impact systems on solids, contact angle hysteresis filters impact energy into modal interfacial vibrations. 
\\
The continuity equation for incompressible flow inside the impacting droplet is given as
\begin{equation}
    {\nabla}{\cdot}\mathbf{V}=0
\end{equation}
The velocity field $\mathbf{V}$ inside the droplet could be extracted from a velocity potential ${\Phi}$ provided that the flow inside the impacting droplet is inviscid. This could be done due to the fact that the impacting droplet is about $350$ times less viscous than the silicone oil pool. The experimental response time scale of the droplet is within the capillary time scale ($t_c{\sim}\mathcal{O}(9{\times}10^{-3}s)$). The viscous effects inside the droplet are important at the scale of the viscous diffusion time scale (${\sim}\mathcal{O}(4R_0^2/{\nu}_w)=5.43s$). We should realize that the droplet response occurs at a scale of milliseconds which is three orders of magnitude smaller than the viscous diffusion time scale inside the droplet. We therefore use a inviscid droplet model for computing the droplet response during the capillary time scale.  
\begin{equation}
    \mathbf{V}=\mathbf{\nabla}{\Phi}
\end{equation}
Substituting equation (3.24) in equation (3.23) we have
\begin{equation}
    {\nabla}^2{\Phi}=0
\end{equation}
where ${\nabla}^2$ is the laplacian operator. We use spherical polar coordinates $(r,{\theta},{\Psi})$ where $r$ represents the radial coordinate from the centre of the droplet, ${\theta}$ is the zenith/polar angle and ${\Psi}$ is the azimuthal angle. We assume an axisymmetric geometry
due to a cylindrical symmetry about the y-axis (Fig. 10(a)) in our experimental configuration.  
Note the local analysis carried out in the previous section uses a different nomenclature for various angular coordinates ${\theta}$ and ${\psi}$ (refer Fig. 3(c)). 
Equation (3.25) in spherical axisymmetric polar coordinates therefore becomes
\begin{equation}
    {\nabla}^2{\Phi}=sin{\theta}(r^2{\phi}_r)_r+(sin{\theta}{\Phi}_{\theta})_{\theta}=0
\end{equation}
on 
\begin{equation}
    \mathscr{D}^i{\equiv}\{(r,{\theta}){\:}|{\:}0{\leq}r{\leq}R{\:},{\:}0{\leq}{\theta}{\leq}{\pi}\}
\end{equation}

\begin{equation}
    \mathscr{D}^e{\equiv}\{(r,{\theta}){\:}|{\:}R{\leq}r{\leq}{\infty}{\:},{\:}0{\leq}{\theta}{\leq}{\pi}\}
\end{equation}

We use subscript $r$, ${\theta}$ in equation (3.26) to represent partial derivatives with respect to radial coordinate ($r$) and the zenith angle (${\theta}$) respectively. This convention is used throughout the analysis in this section. $\mathscr{D}$ represents the 2D axisymmetric equilibrium domain of the problem. The superscripts $i$, $e$ represents regions pertaining to the interior and exterior of the droplet equilibrium shape respectively. $\mathscr{D}^i$ in equation (3.27) represents the 2D axisymmetric interior region of the equilibrium shape of the impacting droplet. 
Similarly the exterior of the droplet in equilibrium conditions is given by $\mathscr{D}^e$ represented in equation (3.28). The pressure inside and outside the droplet could be computed from the unsteady Bernoulli equation \cite{kundu2015fluid}
\begin{equation}
    p(r,{\theta},t)=p_0+{\rho}{\Phi}_t
\end{equation}
where $p_0$ is a reference pressure (the atmospheric pressure), and ${\rho}$ is the density of the fluid. The subscript $t$ refers to partial derivative of ${\Phi}$ with respect to time.
The droplet interface is represented by ${\partial}\mathscr{D}$. The superscript $s$ will represent the submerged interface (water-silicone oil interface) and the superscript $f$ represents the free surface of the droplet (air-water interface).
On the submerged boundary portion of the droplet represented by
\begin{equation}
    {\partial}\mathscr{D}^s{\equiv}\{(r,{\theta}){\:}|{\:}r=R{\:},{\:}{\pi}-{\alpha}{\leq}{\theta}{\leq}{\pi}\}
\end{equation}
the radial velocity ${\Phi}_r$ and the amplitude $Z$ is zero. This approximation could be made due to the fact that the liquid pool is highly viscous and dissipates any interface oscillatory motion. The time scale of the viscous dissipation in the liquid pool is of the order of ${\sim}\mathcal{O}(4R_0^2/{\nu}_s){\sim}10{\times}10^{-3}s{\sim}t_c$
\begin{equation}
    {\Phi}_r = 0{\:};{\:}Z=0
\end{equation}

On the free boundary portion of the droplet
\begin{equation}
    {\partial}\mathscr{D}^f{\equiv}\{(r,{\theta}){\:}|{\:}r=R{\:},{\:}0{\leq}{\theta}{\leq}{\pi}-{\alpha}\}
\end{equation}
the boundary condition at the interface preserving the continuity of normal velocity is given by
\begin{equation}
    {\Phi}_r = -Z_t
\end{equation}
The dynamic balance of momentum along the droplet radius gives
\begin{equation}
    p^i - p^e = {\sigma}\left[\frac{2}{R}-\frac{1}{R^2}\left(\frac{1}{sin{\theta}}\left(sin{\theta}Z_{\theta}\right)_{\theta}+2Z\right)\right]
\end{equation}
Recognize that the unknowns are the droplet deformation from equilibrium position given by $Z({\theta},t)$ and the velocity potential given by ${\Phi}(r,{\theta},t)$
\begin{equation}
    Z = Z({\theta},t)={\:}?{\:};{\:}{\Phi}(r,{\theta},t)={\:}?
\end{equation}
Conservation of mass for the impacting drop gives us
\begin{equation}
    \int_0^{\pi}Zsin{\theta}d{\theta}=0
\end{equation}
This is true if the droplet does not fragment into daughter droplets which is true at low impact energies as in the current experimental configuration.
By considering periodic motion inside and of the free air-water interface, we can set
\begin{equation}
    {\Phi}(r,{\theta},t)={\phi}(r,{\theta})e^{i{\omega}t}
\end{equation}

\begin{equation}
    Z({\theta},t) = z(\theta)e^{i({\omega}t+{\pi}/2)}
\end{equation}
The phase difference of ${\pi}/2$ is consistent with the boundary condition given by equation (3.33).
The partial derivative of ${\Phi}$ with respect to the radial coordinate $r$ is given by
\begin{equation}
    {\Phi}_r = {\phi}_re^{i{\omega}t}
\end{equation}
The partial derivative of $Z$ with respect to time $t$ is given by
\begin{equation}
    Z_t = z({\theta})i{\omega}e^{i{\omega}t}e^{i{\pi}/2} = -{\omega}z({\theta})e^{i{\omega}t}
\end{equation}

Using equation (3.39) and (3.40) in (3.33) we have
\begin{equation}
    {\phi}_re^{i{\omega}t}={\omega}z({\theta})e^{i{\omega}t}
\end{equation}
The pressure inside the droplet $p^i$ using equation (3.29) is given by
\begin{equation}
    p^i = p_0^i+{\rho}^i{\Phi}^i_t
\end{equation}
where ${\rho}^i$ is density of the water droplet.
The pressure outside the droplet $p^e$ using equation (3.29) is given by
\begin{equation}
    p^e = p_0^e+{\rho}^e{\Phi}^e_t
\end{equation}
where ${\rho}^e$ is density of the surrounding air.
The difference in pressure inside and outside of the droplet across the air-water interface is given by
\begin{equation}
    p^i-p^e = (p_0^i-p_0^e) + {\rho}^i{\Phi}^i_t - {\rho}^e{\Phi}^e_t
\end{equation}

The partial derivative of equation (3.37) with respect to time can be written as
\begin{equation}
    {\Phi}_t = i{\omega}{\phi}(r,{\theta})e^{i{\omega}t}
\end{equation}
Using equation (3.45) in equation (3.44) for both internal and external fluids we have
\begin{equation}
    p^i-p^e = (p_0^i-p_0^e) + i{\omega}e^{i{\omega}t}[{\rho}^i{\phi}^i-{\rho}^e{\phi}^e]
\end{equation}
Equation (3.46) can be simplified using the Euler's identity $e^{i{\omega}t}=cos({\omega}t)+isin({\omega}t)$ where $i$ is the complex number ($i=\sqrt{-1}$)
\begin{equation}
    p^i-p^e = (p_0^i-p_0^e) - {\omega}sin{\omega}t[{\rho}^i{\phi}^i-{\rho}^e{\phi}^e]+i{\omega}cos{\omega}t[{\rho}^i{\phi}^i-{\rho}^e{\phi}^e]
\end{equation}
The partial derivative of equation (3.38) with respect to ${\theta}$ can be written as
\begin{equation}
    Z_{\theta}=iz^{'}({\theta})e^{i{\omega}t}
\end{equation}
Using equation (3.48) the curvature term in the right hand side of equation (3.34) can be written as
\begin{equation}
\left[\frac{2}{R}-\frac{1}{R^2}\left(\frac{1}{sin{\theta}}\left(sin{\theta}Z_{\theta}\right)_{\theta}+2Z\right)\right]=I + II
\end{equation}

\begin{equation}
    I = -\frac{1}{R^2}\left[cot{\theta}z^{'}({\theta})+z^{''}({\theta})+2z({\theta})\right]sin{\omega}t
\end{equation}

\begin{equation}
    II = \frac{i}{R^2}\left[cot{\theta}z^{'}({\theta})+z^{''}({\theta})+2z({\theta})\right]cos{\omega}t
\end{equation}

Using equation (3.47), (3.50) and (3.51) in equation (3.34) and equating real parts we have
\begin{equation}
    p^i_0 - p^e_0 = \frac{2{\sigma}}{R}
\end{equation}
and from equating imaginary parts we have
\begin{equation}
    - {\omega}[{\rho}^i{\phi}^i-{\rho}^e{\phi}^e]=\frac{\sigma}{R^2}\left[cot{\theta}z^{'}({\theta})+z^{''}({\theta})+2z({\theta})\right]
\end{equation}
Equation (3.52) reresents the laplace equation for the unperturbed spherical droplet.
On the 2D axisymmetric domain $\mathscr{D}$ we have
\begin{equation}
    {\nabla}^2{\phi}=0 
\end{equation}
and from equation (3.41) we have
\begin{equation}
    {\phi}_r={\omega}z({\theta})
\end{equation}
On the free air-water interface of the droplet ${\partial}\mathscr{D}^f$ we have from equation (3.53)
\begin{equation}
    {\omega}[{\rho}^i{\phi}^i-{\rho}^e{\phi}^e]=-\frac{\sigma}{R^2}\left[\frac{1}{sin{\theta}}\left(sin{\theta}Z_{\theta}\right)_{\theta}+2Z\right]
\end{equation}
On the submerged water-oil interface of the droplet ${\partial}\mathscr{D}^s$ we have from equation (3.31), (3.33), (3.37) and (3.38)
\begin{equation}
    {\phi}_r=0{\:},{\:}z=0
\end{equation}
Using equation (3.38) in equation (3.36) we have
\begin{equation}
    \int_0^{\pi}zsin{\theta}d{\theta}=0
\end{equation}
Equation (3.54) - equation (3.58) defines an eigen value problem in a continuous functional space.
 The equation
\begin{equation}
    {\nabla}^2{\phi}=0
\end{equation}
on $\mathscr{D}$ and the equation
\begin{equation}
    {\phi}_r={\omega}z
\end{equation}
on ${\partial}\mathscr{D}$ defines an Neumann problem. Further due to stagnant nature of the ambient at far field $r{\rightarrow}{\infty}$ we have 
\begin{equation}
    {\phi}^e{\rightarrow}0{\:}as{\:}r{\rightarrow}{\infty}
\end{equation}
Introducing the change of variable ${\mu}=cos{\theta}$, the velocity potential solution for the interior and exterior of the droplet could be written as infinite series using the Legendre Polynomials as the basis functions. The velocity potential therefore could be written as
\begin{equation}
    {\phi}^i(r,{\mu})={\omega}R\left({\phi}_0+\sum_{k=1}^{\infty}\frac{{\phi}_k}{k}\frac{r^k}{R^k}P_k({\mu})\right)
\end{equation}

\begin{equation}
    {\phi}^e(r,{\mu})=-{\omega}R\sum_{k=1}^{\infty}\frac{{\phi}_k}{k+1}\frac{R^{k+1}}{r^{k+1}}P_k({\mu})
\end{equation}
where
\begin{equation}
    {\phi}_k=\frac{<z,P_k>}{<P_k,P_k>}
\end{equation}

\begin{equation}
    <f,g>=\int_{-1}^{+1}fgd{\mu}
\end{equation}

\begin{equation}
    <P_k,P_k>=\frac{2}{2k+1}
\end{equation}
Here $P_k({\mu})$ denotes the k-th Legendre polynomial and $<f,g>$ represents the inner product of $f$ and $g$ and is square integragble in the domain ($L^2(-1,1)$). Equation (3.56) using the definition of ${\mu}$ could be written as
\begin{equation}
    [(1-{\mu}^2)z{\mu}]_{\mu}+2z=-\frac{{\rho}^i{\omega}R^2}{\sigma}\left[{\phi}^i(R,{\mu}) - \frac{{\rho}^e}{{\rho}^i}{\phi}^e(R,{\mu})\right]=f({\mu})
\end{equation}

for $a{\leq}{\mu}{\leq}1$. Here $a=-cos({\alpha})$ (refer Fig. 10(a)).

If $f({\mu})=0$, equation (3.67) has two independent solutions

\begin{equation}
    P_1({\mu})={\mu}{\:},{\:}Q_1({\mu})=\frac{1}{2}{\mu}log{\left(\frac{1+{\mu}}{1-{\mu}}\right)} - 1
\end{equation}

The solution of equation (3.67) may be expressed as
\begin{equation}
    z({\mu})=\int_a^{1}G^{'}({\mu},{\tau})f({\tau})d{\tau}
\end{equation}

where the symmetric Green function is given by the following mapping
\begin{equation}
    G^{'}({\mu},{\tau}):[a,1]{\times}[a,1]{\rightarrow}R
\end{equation}

\begin{equation}
     G^{'}({\mu},{\tau})=\left\{
     \begin{array}{ll}
        P_1({\tau})\left[\frac{Q_1(a)}{a}P_1({\mu}) - \frac{P_1(a)}{a}Q_1({\mu})\right]{\:},{\:}a{\leq}{\mu}{\leq}{\tau}{\leq}1    \\\\
        P_1({\tau})\left[\frac{Q_1(a)}{a}P_1({\mu}) - \frac{P_1(a)}{a}Q_1({\mu})\right]{\:},{\:}a{\leq}{\mu}{\leq}{\tau}{\leq}1  
     \end{array}
     \right.
\end{equation}

When $a{\rightarrow}0$ or $a{\rightarrow}-1$, the green function diverges.
\begin{equation}
    \lim_{{\mu} \to a^+} z({\mu}) = 0{\:},{\:} -1{\leq}{\mu}{\leq}a
\end{equation}
Equation (3.57) gives $z({\mu})=0$ on ${\partial}\mathscr{D}^s$.
Therefore for the entire boundary of the droplet ${\partial}\mathscr{D}={\partial}\mathscr{D}^f{\cup}{\partial}\mathscr{D}^s$ we have 
\begin{equation}
    z({\mu})=\int_{-1}^{+1}G({\mu},{\tau})f({\tau})d{\tau}
\end{equation}
where $G({\mu},{\tau}):[-1,1]{\times}[-1,1]{\rightarrow}R$ is the extension of $G^{'}({\mu},{\tau})$ for $({\mu},{\tau}){\in}[a,1]{\times}[a,1]$

\begin{equation}
    G({\mu},{\tau})=\left\{
        \begin{array}{ll}
             G^{'}({\mu},{\tau}){\:},{\:}({\mu},{\tau})\in[a,1]{\times}[a,1]\\
             0{\:},{\:}otherwise
        \end{array}
    \right.
\end{equation}
From equations (3.62) and (3.63), we may evaluate the function
$f({\tau})$ appearing in equation (3.73) as
\begin{equation}
    f({\mu})=-\frac{1}{\lambda}\left[{\phi}_0+\int_{-1}^{+1}{\Gamma}({\mu},{\tau})z({\tau})d{\tau}\right]
\end{equation}
where 
\begin{equation}
    {\lambda}=\frac{\sigma}{{\rho}^i{\omega}^2R^3}
\end{equation}
and 
\begin{equation}
    {\Gamma}({\mu},{\tau})=\sum_{k=1}^{\infty}\frac{2k+1}{2}\left[\frac{1}{k}+\frac{{\rho}^e}{{\rho}^i}\frac{1}{k+1}\right]P_k({\mu})P_k({\tau})
\end{equation}
Equation (3.73) after substitution of equation (3.75) becomes
\begin{equation}
    z({\mu})=-\frac{1}{\lambda}\left[{\phi}_0\int_{-1}^{+1}G({\mu},{\tau})d{\tau}+\int_{-1}^{+1}\int_{-1}^{+1}G({\mu},{\tau}){\Gamma}({\tau},{\sigma})z({\sigma})d{\sigma}d{\tau}\right]
\end{equation}
The constant ${\phi}_0$ can be determined from equation (3.58) by imposing the conservation of droplet mass/volume.
\begin{equation}
    {\phi}_0 = - \frac{\int_{-1}^{+1}\int_{-1}^{+1}\int_{-1}^{+1}G({\nu},{\tau}){\Gamma}({\tau},{\sigma})d{\nu}d{\sigma}d{\tau}}{\int_{-1}^{+1}\int_{-1}^{+1}G({\nu},{\tau})d{\nu}d{\tau}}
\end{equation}
Using the value of ${\phi}_0$ from equation (3.79) equation (3.78) becomes
\begin{equation}
    {\lambda}z({\mu})=\int_{-1}^{+1}K({\mu},{\sigma})z({\sigma})d{\sigma}=\mathscr{A}z{\:};{\:}(-1{\leq}{\mu}{\leq}1)
\end{equation}
where

\begin{equation}
    K({\mu},{\sigma})=\frac{\int_{-1}^{+1}\int_{-1}^{+1}\int_{-1}^{+1}G({\mu},{\epsilon})G({\nu},{\tau}){\Gamma}({\tau},{\sigma})d{\tau}d{\nu}d{\epsilon}}{\int_{-1}^{+1}\int_{-1}^{+1}G({\nu},{\tau})d{\tau}d{\nu}}-\int_{-1}^{+1}G({\mu},{\tau}){\Gamma}({\tau},{\sigma})d{\tau}
\end{equation}

\begin{figure}
  \centerline{\includegraphics[scale=1]{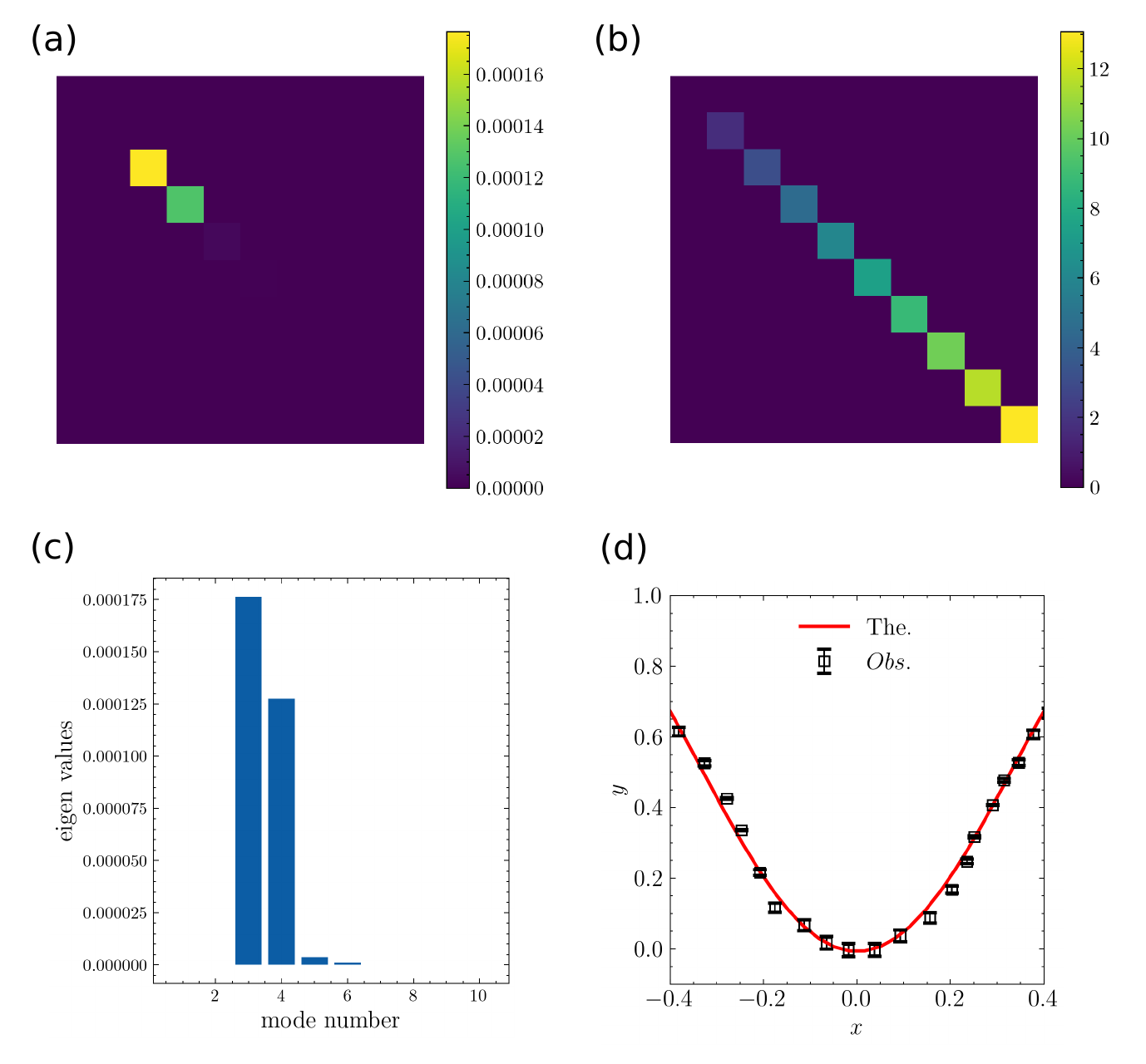}}
  \caption{(a) Visual representation of matrix $A$ depicting its elements.
  (b) Visual representation of the eigenvector matrix of $A$. (c) Bar plot depicting the strength (eigenvalues) of various eigenmodes of matrix $A$. (d) Experimental and theoretical comparison of the the air-water interface crater profile for impact Weber number $We=16$. 
  }
\label{Figure11}
\end{figure}

The continuous functional eigenvalue problem represented by equation (3.80) could be  reduced to an infinite set of algebraic equations by expanding $z({\mu})$ and $K({\mu},{\sigma})$ in the Legendre Polynomial basis as
\begin{equation}
    z({\mu})=\sum_{k=1}^{\infty}z_kP_k(\mu){\:},{\:}z_k=\frac{<z,P_k>}{P_k,P_k}
\end{equation}

\begin{equation}
    K({\mu},{\sigma})=\sum_{h,l=1}^{\infty}K_{hl}P_h({\mu})P_l({\sigma})
\end{equation}

\begin{equation}
    G({\mu},{\tau})=\sum_{h,l=0}^{\infty}G_{hl}P_h({\mu})P_l({\sigma})
\end{equation}
Using equation (3.74) we have
\begin{equation}
    G_{hl}=G_{lh}=\frac{<<G,P_h>,P_l>}{<P_h,P_h><P_l,P_l>}=\frac{(2h+1)(2l+1)}{4}\int_a^1\int_a^1G^{'}({\mu},{\sigma})P_h({\mu})P_l({\sigma})d{\mu}d{\sigma}
\end{equation}

Substituting equation (3.84), (3.85) for $G({\mu},{\tau})$ and equation (3.77) for ${\Gamma}({\tau},{\sigma})$ in equation (3.81) we have
\begin{equation}
    K_{hl}=\left(\frac{G_{0l}G_{h0}}{G_{00}}-G_{lh}\right)\left(\frac{1}{l}+\frac{{\rho}^e}{{\rho}^i}\frac{1}{l+1}\right)
\end{equation}
using equation (3.82) and (3.83) in (3.80) we have
\begin{equation}
    {\lambda}z_h=\sum_{l=1}^{\infty}\left[\frac{2}{2l+1}\left(\frac{G_{h0}G_{0l}}{G_{00}}-G_{hl}\right)\left(\frac{1}{l}+\frac{1}{l+1}\right)\right]
\end{equation}
Using a linear coordinate transformation between $x_h$ and $z_h$ we have
\begin{equation}
    x_h=\left(\frac{\frac{1}{h}+\frac{{\rho}^e}{{\rho}^i}\frac{1}{h+1}}{2h+1}\right)^{1/2}z_h
\end{equation}

\begin{equation}
   \sum_{l=1}^{\infty}A_{lh}x_l= {\lambda}x_h
\end{equation}

Equation (3.89) describes an infinite dimensional matrix-vector eigenvalue problem where $A_{lh}$ represents the matrix, $x_l$ or $x_h$ represents the eigenvector and ${\lambda}$ represents the eigenvalue corresponding to the eigenvector $x_h$.
The elements of the matrix $A$ is given by
\begin{equation}
    A_{lh}=A_{hl}=2\left(\frac{G_{h0}G_{0l}}{G_{00}}-G_{hl}\right)\left[\left(\frac{\frac{1}{h}+\frac{{\rho}^e}{{\rho}^i}\frac{1}{h+1}}{2h+1}\right)\left(\frac{\frac{1}{l}+\frac{{\rho}^e}{{\rho}^i}\frac{1}{l+1}}{2l+1}\right)\right]
\end{equation}
 Equation (3.90) depicts that the matrix $A$ is real and symmetric.
\begin{equation} 
G_{ik}= \frac{(2i+1)(2k+1)}{4}\left\{A_iB_k+C_{ik}\right\}
\end{equation}
\begin{equation}
    A_i=\int_a^1P_1({\tau})P_i({\tau})d{\tau}
\end{equation}
\begin{equation}
    B_k = \int_a^1\left(\frac{Q_1(a)}{a}P_1({\mu})-\frac{P_1(a)}{a}Q_1({\mu})\right)P_k({\mu})d{\mu}
\end{equation}

\begin{equation}
    C_{ik}=\frac{P_1(a)}{a}\int_{a}^1[P_i({\tau})P_1({\tau})\int_{\tau}^1Q_1({\mu})P_k({\mu})d{\mu}-Q_1({\tau})P_i({\tau})\int_{\tau}^1P_1({\mu})P_k({\mu})d{\mu}]d{\tau}
\end{equation}

Equation (3.94) could be simplified based on the identities \citep{macrobert1967spherical} for $k{\neq}1$
\begin{equation}
    \int_x^1P_h({\mu})P_k({\mu})d{\mu} = \frac{hP_k(x)P_{h-1}(x)-kP_h(x)P_{k-1}(x)-(h-k)xP_h(x)P_x(x)}{(h-k)(h+k+1)}
\end{equation}

\begin{equation}
    \int_x^1Q_h({\mu})P_k({\mu})d{\mu}=\frac{hP_k(x)Q_{h-1}(x)-kQ_h(x)Q_{k-1}(x)-(h-k)xQ_h(x)P_k(x)}{(h-k)(h+k+1)}
\end{equation}

If $k{\neq}1$
\begin{equation}
    G_{ki}=G_{ik}=\frac{(2i+1)(2k+1)}{4}\left[\frac{P_k(a)I_{1i}(a)-P_1(a)I_{ik}(a)}{a[k(k+1)-2]}\right]
\end{equation}
where 
\begin{equation}
    I_{ik}(a)=\int_a^1P_i({\tau})P_k({\tau})d{\tau}
\end{equation}
and
\begin{equation}
    I_{1i}(a)=I_{ik}(a)|_{i=1,k=i}
\end{equation}

For $i=k=1$
\begin{equation}
    G_{11}=\frac{1}{2}log(1+a)-\frac{1}{4a}+\frac{7-6log(2)}{12}-\frac{a^3}{12}-\frac{a}{4}
\end{equation}

The symmetric Green function $G^{'}$ represented by equation (3.71) diverges for $a=-1$ and $a=0$. The eigenvalue problem described by equation (3.54) to equation (3.58) cannot be solved in the function space $L^2(-1,1)$. The function space used should be different \citep{benjamin1981trends} such that $\lim_{a \to 0}A_{hk}$ exists and is finite. The singularity for $a=0$ could be easily resolved by setting
\begin{equation}
    \mathscr{A}_{a=0}=\lim_{a \to 0}\mathscr{A}_a
\end{equation}
Further, the limiting value of the operator $\mathscr{A}$ and its eigenvalues as ${a}{\rightarrow}0$ and $a{\rightarrow}-1^+$ is shown below.
\begin{equation}
    \lim_{a \to 0} A_{hk} = 2\left(\frac{\frac{1}{h}+\frac{{\rho}^e}{{\rho}^i}\frac{1}{h+1}}{2h+1}\frac{\frac{1}{k}+\frac{{\rho}^e}{{\rho}^i}\frac{1}{k+1}}{2k+1}\right)^{1/2}
\end{equation}
From equation (3.102) we observe that $\lim_{{\alpha} \to 0}A_{hk}$ exists and is finite.

\begin{equation}
    \lim_{a \to 0}\left(\frac{G_{h0}G_{k0}}{G_{00}}-G_{hk}\right)=\lim_{a \to 0}\left[\frac{1}{a}\left(\frac{aG_{h0}aG_{k0}}{aG_{00}}-aG_{hk}\right)\right]
\end{equation}

\begin{equation}
    aG_{hk}=\frac{(2h+1)(2k+1)}{4}\left[\frac{P_k(0)I_{1n}(0)}{k(k+1)-2}+\frac{P_k(0)I_{1h}(0)-I_{hk}(0)}{k(k+1)-2}a+\mathcal{O}(a^2)\right]
\end{equation}
For $k{\neq}1$,
\begin{equation}
    aG_{11}=\frac{9}{4}\left[-\frac{1}{9}+\frac{1}{27}(7-6log2)a+\mathcal{O}(a^2)\right]
\end{equation}

\begin{equation}
    \left(\frac{aG_{h0}aG_{k0}}{aG_{00}}-aG_{kk}\right)_{a=0}=0{\:};{\:}{\forall}h,K
\end{equation}
The derivative of equation (3.106) at $a=0$ is finite, thus showing the existence of a finite value of the limit depicted in equation (3.103).

\begin{equation}
    \lim_{a \to 0}\left(\frac{G_{h0}G_{k0}}{G_{00}}-G_{hk}\right)=\frac{(2h+1)(2k+1)}{4[k(k+1)-2]}\left[P_k(0)\left(2I_{1h}(0)-\int_0^1P_h({\tau})d{\tau}\right)-E\right]
\end{equation}

\begin{equation}
    E = 2I_{1h}(0)\int_0^1P_k({\tau})d{\tau} + I_{hk}(0)
\end{equation}

\begin{equation}
    \lim_{a \to -1^+}G_{ik}=\frac{(2i+1)(2k+1)}{4}\frac{(-1)^k\frac{2{\delta}_{i1}}{3}+\frac{2{\delta}_{ik}}{2k+1}}{2-k(k+1)}
\end{equation}
For $k{\neq}1$ the matrix elements of $A$ becomes
\begin{equation}
    A_{1k}=A_{k1}{\rightarrow}\frac{(-1)^k}{k(k+1)-2}\left[\left(1+\frac{1}{2}\frac{{\rho}^e}{{\rho}^i}\right)\left(\frac{1}{k}+\frac{1}{k+1}\frac{{\rho}^e}{{\rho}^i}\right)\left(\frac{2k+1}{3}\right)\right]^{1/2}
\end{equation}

\begin{equation}
    A_{hk}{\rightarrow}0{\:};{\:}(h{\neq}k,h,k{\neq}1)
\end{equation}
The diagonal elements of $A$ becomes
\begin{equation}
    A_{hh}{\rightarrow}\left(\frac{1}{h}+\frac{{\rho}^e}{{\rho}^i}\frac{1}{h+1}\right)\frac{1}{h(h+1)-2}{\:};{\:}(h{\neq}1)
\end{equation}
Equations (3.110, (3.111) and (3.112) depicts that $A$ is a symmetric real diagonal matrix with eigenvalues as the elements of the diagonal.  
The eigenvalues becomes
\begin{equation}
    {\lambda}^1{\rightarrow}{+\infty}{\:},{\:}x_k^1{\rightarrow}{\delta}_{1k}{\:},{\:}z_k^1{\rightarrow}{\delta}_{1k}
\end{equation}

\begin{equation}
    {\lambda}^n{\rightarrow}\frac{(n+1)+({\rho}^e/{\rho}^i)n}{n(n-1)(n+1)(n+2)}{\:};{\:}x_k^n{\rightarrow}{\delta}_{nk}{\:},{\:}z_k^n{\rightarrow}{\delta}_{nk}
\end{equation}
From equations (3.113) and (3.114) we observe that the first eigenmode corresponds to zero frequency which corresponds to a rigid body motion, whereas the various other eigenmodes tend towards pure Rayleigh modes for the case of negligible pinning at and below the dynamic contact line.
The infinite dimensional eigenvalue problem represented by equation (3.89) is solved numerically using in-house program written in python using a $10$ dimensional Legendre function space approximation. 
The elements of the matrix A were represented using equation (3.90). Note that the matrix A is real and symmetric. Fig. 11(a) depicts a visual representaion of the matrix $A$. Note that the matrix $A$ is diagonal. Further $A$ being real and symmetric has real eigenvalues and the eigenvalues are the ordered diagonal elements. Matrix $A$ represent a self-adjoint operator in an orthonormal basis of Legendre polynomials forming a real inner product space. Fig. 10(b) depicts the eigenvector matrix for $A$. The columns of the matrix represents each eigenvector while the eigenvector matrix being diagonal represent the orthogonal nature of the eigenvectors in the function space. Fig. 11(c) represent the eigenvalues of the matrix. The eigenvalues correspond to the various modes of the Legendre polynomials used in the approximate function space for calculating the amplitude $z({\theta})$, ${\omega}$, and ${\phi}(r,{\theta})$. Fig. 11(a) and Fig. 11(c) depicts mode $3$, $4$ are the most dominant ones and can approximate the amplitude, frequency and velocity potential functions. Fig. 11(d) shows the comparison of the theoretical curve obtained from the global analysis with the experimental data at the maximum penetration width for impact Weber number of $16$. The experimental and theoretical curves conforms within the experimental error bounds. Further it agrees with the local analysis outlined in the previous section (3.6). For larger impact Weber number, extreme air-crater deformations occur (Fig. 7) and the above analysis needs to be generalized for high amplitudes ${z({\theta})}$. Further, in a general scenario the effects of viscosity has to be incorporated in the shape model and the pinned boundary conditions needs to be applied on the dynamic 1D contact line.
\section{Conclusion}
In conclusion, we unearthed the 
mechanics of air crater formation on the air-water interface during impact on an immiscible viscous liquid pool using high speed imaging and theoretical analysis.  
We found that the penetration process was smooth without the formation air craters/jets for low impact energies. The smooth penetration was due to the delayed air layer rupture time scale ($t_r{\sim}\mathcal{O}(2{\times}10^{-1}s)$). 
For higher impact energies, the air layer rupture time scale is reduced by three orders of magnitude ($t_r{\sim}\mathcal{O}(5{\times}10^{-4}s)$). The droplet experiences an impulsive force that results in the formation of capillary waves, 
causing the formation of air craters.
The major cause of the air crater formation is the sudden deceleration of the impinging droplet.
The retardation of the droplet is primarily due to viscous forces and is a monotonically increasing function of impact Weber number. We discovered that significant depth air craters during penetration are formed when the ratio of viscous force and capillary force exceeds unity (${\Gamma}={{\Delta}p_s}/{{\Delta}p_c}>1$). Further, the air crater response time scale is also a monotonically increasing function of Weber number ($T{\sim}We^{1/2}$). 
Further, in general we found that the air crater/jet profiles could be described by the superposition of dominant eigenmodes in Legendre polynomial basis. The air crater/jet profiles were computed using a local and global analysis based on the eigenvalue problem of a inviscid constrained Rayleigh drop model with a dynamic contact line. Due to high viscosity, the entire drop surface below and including the dynamic contact line had negligible amplitude and hence assumed to be pinned. The air craters observed in the experiment were caused due to the interfacial deformation of the droplet surface area above the dynamic contact line. In future, we will generalize the idea of droplet surface response during imapct on immiscible liquids by considering the viscous Rayleigh drop model and allowing zero amplitude (pinned contact) only on the 1D dynamic contact line rather than on a 2D surface. 


\section*{Supplementary movie captions}
Movie 1: Supplementary movie for impact Weber number ($We=4$). Recorded at 10000 FPS, playback speed 100FPS.

Movie 2: Supplementary movie for impact Weber number ($We=16$). Recorded at 10000 FPS, playback speed 10FPS.

Movie 3: Supplementary movie for impact Weber number ($We=145$). Recorded at 10000 FPS, playback speed 10FPS.

\section*{Acknowledgement}
The authors are thankful for the funding received from the Defence Research and Development Organization (DRDO) Chair Professorship.

\section*{Declaration of Interests}
The authors declare no conflict of interest.
\clearpage
\bibliographystyle{jfm}
\bibliography{jfm-instructions}

\end{document}